\documentclass[fleqn,usenatbib]{mnras}

\usepackage{newtxtext,newtxmath}

\usepackage[T1]{fontenc}

\DeclareRobustCommand{\VAN}[3]{#2}
\let\VANthebibliography\thebibliography
\def\thebibliography{\DeclareRobustCommand{\VAN}[3]{##3}\VANthebibliography}

\usepackage{graphicx}	
\usepackage{amsmath}	
\usepackage{tikz}

\newcommand{\cii}{{[}C\,\textsc{ii}{]}}
\newcommand{\hi}{H\,\textsc{i}}
\newcommand{\avg}[1]{\left\langle#1\right\rangle}

\usepackage{color}

\newif\iftrack
\iftrack
\newcommand{\added}[1]{{\bf #1}}
\newcommand{\deleted}[1]{}
\newcommand{\replaced}[2]{{\bf #2}}
\else
\newcommand{\added}[1]{{#1}}
\newcommand{\deleted}[1]{}
\newcommand{\replaced}[2]{{#2}}
\fi

\title[\cii{} LIM model and statistics]{The informativeness of \cii{} line-intensity mapping as a probe of the H\textsc{\,i} content and metallicity of galaxies at the end of reionization}

\author[P.~Horlaville et al.]{Patrick Horlaville$^{1,2,3}$\thanks{E-mail: phorlaville24@ubishops.ca}, Dongwoo T. Chung$^{2,4}$, J. Richard Bond$^{2}$, Lichen Liang$^{2}$\\
$^{1}$Department of Physics, McGill University, 3600 rue University, Montreal, QC H3A 2T8, Canada\\
$^{2}$Canadian Institute for Theoretical Astrophysics, University of Toronto, 60 St. George Street, Toronto, ON M5S 3H8, Canada\\
$^{3}$Department of Physics \& Astronomy, Bishop's University, 2600 rue College, Sherbrooke, QC J1M 1Z7, Canada\\
$^{4}$Dunlap Institute for Astronomy and Astrophysics, University of Toronto, 50 St. George Street, Toronto, ON M5S 3H4, Canada\\
}

\date{Accepted XXX. Received YYY; in original form ZZZ}

\pubyear{2024}

\begin{document}
\label{firstpage}
\pagerange{\pageref{firstpage}--\pageref{lastpage}}

\maketitle

\begin{abstract}

Line-intensity mapping (LIM) experiments coming online now will survey fluctuations in aggregate emission in the \cii{} ionized carbon line from galaxies at the end of reionization. Experimental progress must be matched by theoretical reassessments of approaches to modelling and the information content of the signal. We present a new model for the halo--\cii{} connection, building upon results from the FIRE simulations suggesting that gas mass and metallicity most directly determine \cii{} luminosity. Applying our new model to an ensemble of peak-patch halo lightcones, we generate new predictions for the \cii{} LIM signal at $z\gtrsim6$. We expect a baseline 4000-hour LIM survey from the CCAT facility to have the fundamental sensitivity to detect the \cii{} power spectrum at a significance of $\replaced{4}{5}\sigma$ at $z\sim6$, with an extended or successor Stage 2 experiment improving significance to $\replaced{36}{48}\sigma$ at $z\sim6$ and achieving $\replaced{8}{11}\sigma$ at $z\sim7.5$. Cross-correlation through stacking, simulated against a mock narrow-band Lyman-break galaxy survey, would yield a strong detection of the radial profile of cosmological \cii{} emission surrounding star-forming galaxies. We also analyse the role of a few of our model's parameters through the pointwise relative entropy (PRE) of the distribution of \cii{} intensities. While the PRE signature of different model parameters can become degenerate or diminished after factoring in observational distortions, various parameters do imprint themselves differently on the one-point statistics of the intrinsic signal. Further work can pave the way to access this information and distinguish different sources of non-Gaussianity in the \cii{} LIM observation.

\end{abstract}


\section{Introduction}
\label{sec:intro}

As the studies of galaxy formation and cosmology advance, galaxies at high redshift and in particular at the end of the epoch of reionization (EoR) are as important as ever. From the perspective of cosmology, understanding how the Universe exits the \deleted{epoch of }EoR relies on understanding the physical conditions, abundances, and cosmic structure of the first stars and galaxies that carry out the reionization process. From the perspective of galaxy formation, to understand the nature of these first galaxies and proto-galaxies is to understand the starting point of cosmic star-formation history, with star-formation activity continuing to rise all the way through the `cosmic noon' of $z\sim2$--3 (see, e.g., the review of~\citealt{MD14}). The EoR thus sits at a key intersection of cosmology and extragalactic astrophysics, between structure formation and galaxy formation. Studying the population of galaxies from the EoR is fundamental to understanding the Universe both as we know it today and as it was almost fourteen billion years ago.

Although targeted or pencil-beam surveys are already detecting individual objects from the EoR, conventional spectroscopic galaxy surveys will struggle to probe EoR galaxies at the efficiency required for this population to act as a true large-scale cosmological probe. Thus, a full understanding of the EoR demands alternative ways to measure the clustering of star formation and gas at high redshift. 

Intensity mapping (IM), or line intensity mapping (LIM), is a technique analysing low-to-medium resolution spatial-spectral observations to probe a three-dimensional cosmological volume for the signature of large-scale structure as traced by spectral lines associated with specific atomic or molecular species. The strength of such analyses resides in their potential to probe large cosmological volume efficiently while measuring the contribution of all galaxies, including those impractical for standard galaxy surveys to detect individually (see reviews by, e.g.: \citealt{Kovetz17,Kovetz19,Bernal22}).

\begin{figure}
    \centering
    \includegraphics[width=0.98\linewidth]{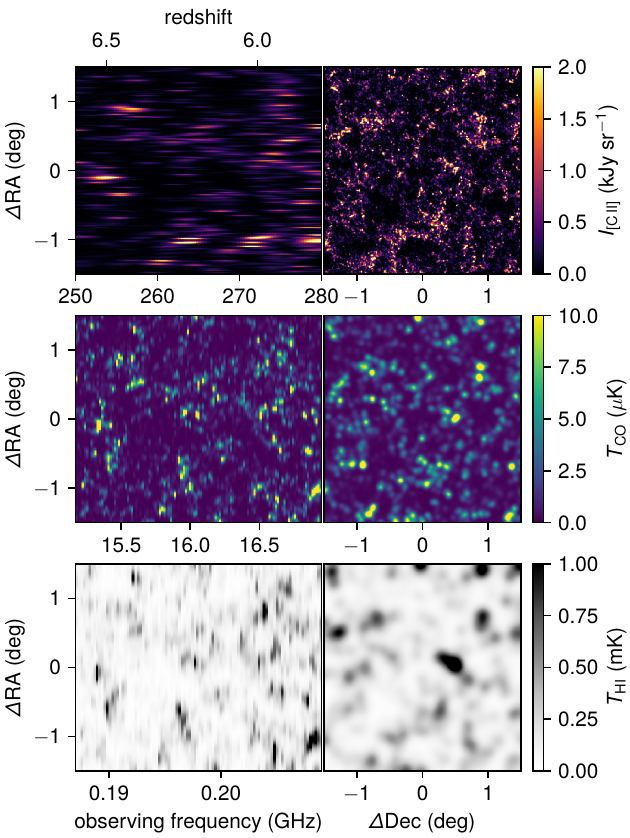}
    \caption{Visualization of correlated LIM observables derived from applying our state-of-the-art understanding of the galaxy--halo connection to one of the cosmological simulations used in this work, and convolved with realistic instrumental responses. The \cii{} signal (top panels), tracing diffuse neutral gas in the ISM, is simulated using the model put forward in this paper, and convolved with a Gaussian angular beam of width $\approx50''$ and a frequency-space profile of width $\sim3$\,GHz, corresponding to the near-future capabilities of the CCAT facility. The CO signal (middle panels), tracing molecular gas, is simulated using a model from prior literature~\citep{Li_2016} and convolved with an angular beam of width $\sim4'$, corresponding to a 10 metre dish as used by the COMAP Pathfinder~\citep{COMAPESI}. The 21\,cm signal as shown neglects contributions outside of halos -- although the intergalactic medium should not dominate the signal, given the high ionized fraction expected at $z\sim6$ -- and is convolved with a beam of width $\sim10'$, corresponding to an interferometer with $\sim500$\,m baselines.}
    \label{fig:prettyviz}
\end{figure}

While atomic hydrogen will be most abundant particularly at early cosmic epochs, the past decade has seen studies of LIM concepts leveraging other spectral lines associated with dust-obscured star-formation activity to complement IM surveys of the 21 cm atomic hydrogen line (see~\autoref{fig:prettyviz}). This includes the rotational transitions of carbon monoxide (CO) observable at centimetre wavelengths, one of the most abundant luminous tracers of molecular gas and thus the subject of significant work in LIM modelling~\citep{Righi08,VL10,Lidz11,Pullen13,Li_2016,Sun19,MoradinezhadKeating19,Yang21,Yang22} and experiment~\citep{Keating16,mmIME-ACA,Keenan22,COMAPESI}. 

At shorter wavelengths, the 157.7 micron \cii{} line from ionized carbon (with rest-frame frequency $\nu_\text{rest}=1900.5$\,GHz) becomes one of the most promising tracers of high-redshift galaxies. In our local volume, \cii{} is a key far-infrared (FIR) cooling line that makes up as much as 1\% of the bolometric FIR luminosity in local gas-rich and highly luminous galaxies~\citep{Stacey91}. Given mounting observations with the Atacama Large Millimetre/submillimetre Array (ALMA) of individual \cii{} lines from $z\sim4$--8 (see, e.g.:~\citealt{Pentericci16,Knudsen16,Smit18,ALPINELF,Fujimoto21}), the existence of an appreciable population of \cii{} emitters in the late EoR appears certain. The brightness of the cosmological \cii{} emission at this relatively metal-poor epoch remains uncertain, however, and has been the subject of a substantive body of prior modelling work (e.g.:~\citealt{Gong12,Uzgil14,Silva15,Yue15,Serra16,Dumitru18,Chung_2020,Karoumpis22}).

At time of writing, a range of EoR \cii{} pathfinders have begun construction or even operation, including CONCERTO~\citep{CONCERTO,CONCERTO_SPIE}, the Tomographic Ionized-carbon Mapping Experiment (TIME;~\citealt{Crites14,Sun21,TIME2022}), and the Epoch of Reionization Spectrometer (EoR-Spec) instrument on the Cerro Chajnantor Atacama Telescope (CCAT) facility~\citep{CCAT_Prime_Collaboration_2022,EoRSpec}. With all these projects in active development and/or analysis phases, and early science results anticipated within this decade, this work aims to push the modelling of the \cii{} LIM signal forward in two important ways: first by reassessing the nature of the underlying halo--\cii{} connection, and second by assessing the information content of the one-point statistics of the \cii{} intensity fluctuations to be observed.


The model of this work distinguishes itself from many previous models by mediating the halo--\cii{} connection not solely through star-formation rate (SFR), but through the neutral atomic hydrogen (\hi{}) mass and gas-phase metallicity in the ISM of the galaxies in dark matter halos, based on insights from the Feedback In Realistic Environments (FIRE) project examined by~\cite{liang2023}. In formulating this halo model, we reproduce certain trends like the evolution of \cii{} luminosity against FIR luminosity, and arrive at a model consistent with current observational constraints on high-redshift \cii{} abundances.

Applying our halo model to an ensemble of approximate cosmological simulations, we aim to answer the following questions:\begin{itemize}\item What is the detectability of \cii{} emission in near- and far-future mm-wave LIM surveys, in either auto- or cross-correlation? \item How do different model parameters affect the voxel intensity distribution of the line-intensity fluctuations, as represented by the differential relative entropy? \item How do observational effects like the instrument response and thermal noise affect the observability of these distinct signatures in relative entropy?\end{itemize}

The paper is structured as follows. After a summary of context around the halo--\cii{} connection and one-point statistics in~\autoref{sec:context}, we present our new \cii{} model and simulation methods in~\autoref{sec:relat}. We then show our results and forecasts in~\autoref{sec:simres}, and conclude with forward-looking discussion in~\autoref{sec:conclus}.

Unless otherwise stated, we assume base-10 logarithms, and a $\Lambda$CDM cosmology with parameters $\Omega_m = 0.286$, $\Omega_\Lambda = 0.714$, $\Omega_b =0.047$, $H_0=100h$\,km\,s$^{-1}$\,Mpc$^{-1}$ with $h=0.7$, $\sigma_8 =0.82$, and $n_s =0.96$, to maintain consistency with previous simulations used by~\cite{COMAPDDE}. Distances carry an implicit $h^{-1}$ dependence throughout, which propagates through masses (all based on virial halo masses, proportional to $h^{-1}$) and volume densities ($\propto h^3$).
\section{Context in previous work}
\label{sec:context}
\subsection{A FIRE-tested formula: moving away from \texorpdfstring{\cii{}}{[C II]} as a tracer of star formation}
The SFR is a highly attractive lever arm for \cii{} modelling at high redshift. With \cii{} being a dominant FIR cooling line mostly unaffected by dust extinction, \cii{} emission is a natural mechanism to balance heating from newly forming stars. Equally, \cii{} emission is attractive as a proxy for SFR at high redshifts, where a signpost for obscured star-formation activity is needed for a full understanding of cosmic star-formation history.

However, previous work has already cast significant doubt that the connection between \cii{} luminosity and SFR is reliably universal~\citep{HC15,Vallini15,Croxall17}, as other factors like metallicity appear to modulate the observed relation. A full review of such work is well beyond the scope of this paper, and we refer the reader to~\cite{liang2023} for further details. The upshot, however, is that such doubts motivate a re-examination of the nature of the halo--\cii{} connection based on a physically motivated picture of \cii{} emission. Such work is particularly important if we are to leverage synergies between mm-wave LIM and 21 cm tomography at high redshifts~\citep{Padmanabhan23}.

To this end, we make use of insights from the work of \cite{liang2023}, who conducted a comprehensive analysis of the relationship between the \cii{} luminosity and star formation rate (SFR) of simulated star-forming galaxies at $z\sim0$--$8$ generated by the Feedback In Realistic Environments (FIRE) project. The mock galaxy sample includes 507 FIREBox galaxies (including 91 at $z\geq6$), which reside not in zoom-in boxes but in a $(15h^{-1}\,$cMpc$)^3$ cosmological box with prescriptions for hydrodynamics and feedback~\citep{FIREbox}, as well as 10 MassiveFIRE zoom-in simulations (6 of which have snapshots at $z\geq6$; see references in~\citealt{liang2023}). Through post-processing of these simulations to derive the interstellar radiation field and the ionization and excitation states of gas clouds, \cite{liang2023} derived \cii{} luminosities for the FIRE galaxies studied, as well as other properties like the SFR and the bolometric IR luminosity $L_\text{IR}$.

Not only do the FIRE simulations reproduce the empirically observed overall correlations between the \cii{} luminosity, SFR, and $L_\text{IR}$, but \cite{liang2023} also reproduce the empirically observed evolution of the \cii{}--SFR or \cii{}--$L_\text{IR}$ relation with both redshift and $L_\text{IR}$. Of interest is the finding that dense molecular regions contribute negligibly to \cii{} emission in the FIRE simulations, which would imply that the link between \cii{} luminosity and SFR is truly not through directly tracing the cool star-forming molecular gas (as is the case for CO, for instance). Rather, the \cii{} emission primarily coincides with the neutral \hi{} regions between the ionized and molecular gas, where ionized carbon is appreciably present and the gas density is sufficiently high to encourage cooling through \cii{} line emission.

The ultimate result of~\cite{liang2023} was a physical picture of \cii{} emission where four factors modulate the scaling between \cii{} and SFR: the gas depletion timescale, the gas-phase metallicity, the gas density, and the fraction of gas in ionized and neutral regions. This work will leverage this proposed formula to move beyond \cii{} as simply a proxy for SFR, and build a new halo model instead relating \cii{} luminosity to the expected metallicity and neutral hydrogen mass in each halo, using this model to build new forecasts for upcoming \cii{} LIM experiments as will take place on the CCAT facility.

\subsection{The importance of one-point statistics in summarising a highly non-Gaussian signal}

In traditional cosmological contexts like studies of the cosmic microwave background, we are used to the unreasonable effectiveness of Gaussian random fields as a description of cosmological fields. What reflects this is the common use of the power spectrum as the primary summary statistic of observations of density fluctuations, including LIM.

However, structure formation is very much a non-linear process resulting in non-Gaussian structures. In LIM contexts,~\cite{Breysse17} first explicitly proposed the use of the probability distribution function of observed voxel intensities -- i.e., the voxel intensity distribution (VID) -- as a way to capture the highly non-Gaussian information present in line-intensity fluctuations, without relying on higher-order statistics whose estimation and covariance pose significant challenges to detection and interpretation. Work continues on understanding the VID and extending one-point statistics in LIM contexts, on both theoretical and numerical fronts~\citep{Breysse19,Ihle19,SatoPolito22,Breysse23,COMAPDDE}.

It is however the logarithm of the VID, rather than the VID itself, that is most closely related to the information content of the LIM observation. \cite{Lee23} put forward a relevant insight in this context, using one-point statistics to quantify the non-Gaussian information injected into the cosmic infrared background (CIB) by gravitational lensing. In particular, that work put forward quantities related to the relative entropy -- or equivalently to the Kullback--Leibler divergence (KLD;~\citealt{KL51}) -- between the intensity probability distributions of the lensed and unlensed CIB.

It is indeed natural to consider the information content in the VID -- i.e., the probability density function (PDF) $P(I)$ -- as the relative entropy from a model PDF $Q(I)$ to the actual $P(I)$:
\begin{equation}
    S_\text{rel}(P\parallel Q) = -\int dI\,P(I)\,\ln{\frac{P(I)}{Q(I)}}.\label{eq:Srel_def}
\end{equation}
However, \cite{Lee23} considered it equally interesting to examine the integrand, the \emph{pointwise relative entropy} (PRE) per intensity bin (dubbed `weighted relative entropy' in that work). We will consider `per intensity bin' (or `per log-intensity bin' in some cases) to be implicit and often refer simply to
\begin{equation}
    dS_\text{rel}(P\parallel Q) \equiv -P(I)\,\ln{\frac{P(I)}{Q(I)}}.
\end{equation}
(We will often switch between using $dS_\text{rel}$ as shorthand for either $dS_\text{rel}/dI$ as defined above or $dS_\text{rel}/d(\log{I})$; the usage will be clear based on, e.g., the binning of the discrete histogram approximating the underlying continuous $P(I)$.) Note that many works (e.g.,~\citealt{MackayITILA}) define `relative entropy' as the negative of our definition in~\autoref{eq:Srel_def}, as that is equal to the KLD. However, our definition naturally leads to the PRE from $Q$ to $P$ being the Shannon information for $P$ minus that for the `starting' $Q$, weighted by $P$.

Using the PRE -- the log-space difference of one distribution against another, weighted by the true distribution -- provides a fuller description of the effect of a given variable on the intensity PDF. The integrated quantity of relative entropy will quantify the net statistical distance by which that variable pushes the PDF away from an alternative expectation, but the \emph{pointwise} relative entropy can provide full templates of how the difference arises at different ranges of intensity. Thus distinguishing different modes of distributional distortion would disentangle a range of astrophysical and cosmological signatures that can imprint non-Gaussianities in the observable Universe.

Towards the conclusion of this work, rather than explicitly forecasting the detectability of these signatures, we will use the PRE to consider the information content in the VID related to a number of astrophysical model parameters. This in turn will guide future work that will more explicitly probe the detectability of the relevant signatures as well as signatures of primordial non-Gaussianity in the clustering and abundances of \cii{} emitters.

\section{Methods}
\label{sec:relat}

\subsection{Line-luminosity model: a novel halo--\texorpdfstring{\cii{}}{[C II]} connection}
\label{sec:secmodel}

As we reviewed in~\autoref{sec:context}, through the analysis of the relationship between SFR and \cii{} emission in $\sim500$ simulated galaxies, \cite{liang2023} derived a linear scaling relationship for the ratio between SFR and the global \cii{} luminosity $L_\text{\cii{}}$ of a galaxy:
\begin{equation}
    \frac{L_\text{\cii{}}}{\rm{SFR}} \propto f_\text{\cii{}}\overline{Z}_{\rm{gas}}t_{\rm{dep}}\overline{n}_{\rm{gas}},
    \label{lichen_eq}
\end{equation}
where $f_\text{\cii{}}$ is the fraction of the total galactic gas mass contained in \hi{} and H\textsc{\,ii} regions,  $t_{\rm{dep}}\equiv M_{\rm{gas}}/\rm{SFR}$ is the galaxy's gas depletion time scale (equal to the total gas mass divided by the SFR), and $\overline{Z}_{\rm{gas}}$ and $\overline{n}_{\rm{gas}}$ are the average gas-phase metallicity and gas particle number density, respectively.\footnote{Both average quantities are \cii{} brightness-weighted averages. Specifically, the `average' gas density corresponds to the value at the location of the median \cii{} luminosity within the galaxy, and the `average' metallicity is the same but is very similar to the mass-weighted gas metallicity.}

By the end of this section, we will be able to transform this relation into a model of \cii{} luminosity as a function of halo mass and redshift. We first establish \hi{} mass as a primary intermediary property in~\autoref{sec:cii_mhi}, before adding dependence on metallicity in~\autoref{sec:cii_mhiz} and a stochastic step in~\autoref{sec:scatter}.

\subsubsection{$L_\text{\cii{}}$ as a function of \texorpdfstring{\hi{}}{H I} mass}
\label{sec:cii_mhi}


By multiplying both sides of Equation \ref{lichen_eq} by the SFR, we recover a proportionality relationship between $L_\text{\cii{}}$ and $M_{\rm{gas}}$:
\begin{equation}
    L_\text{\cii{}} \propto f_\text{\cii{}}M_\text{gas}\overline{Z}_{\rm{gas}}\overline{n}_{\rm{gas}}.
\end{equation}

For now, we neglect the dependence on metallicity and gas density, and consider the proportionality between $L_\text{\cii{}}$ and $f_\text{\cii{}}M_\text{gas}$. The latter quantity effectively ends up being the gas mass contained in \hi{} and H\textsc{\,ii} regions. The original motivation for this being the relevant quantity in the work of~\cite{liang2023} is that it is entirely acceptable to ignore the contribution of molecular (H$_2$) regions to \cii{} emission, as most of the carbon there is in a neutral rather than ionized state. In particular, however, simulations show that it is chiefly the \hi{} regions that dominate the determination of the \cii{} luminosity. This is because, as~\cite{liang2023} discuss, even the \cii{} emission from H\textsc{\,ii} regions -- which often contribute more to the total \cii{} luminosity than \hi{} regions -- technically originates from the interface of the H\textsc{\,ii} gas with the \hi{}-rich regions.

Based on the above considerations, we approximate the \hi{} gas as the main driver of $f_\text{\cii{}}M_{\rm{gas}}$, such that the \cii{} luminosity scales linearly with the \hi{} gas mass $M_{\rm HI}$, with some proportionality coefficient $A_\text{\cii{}}$:
\begin{equation}
    \frac{L_\text{\cii{}}}{L_\odot} = A_\text{\cii{}}\times\frac{M_{\rm HI}}{M_{\odot}}.
    \label{alpha}
\end{equation}

\cite{paco_2018} examine halos and galaxies in cosmological simulations to provide a relationship between the total halo mass $M_{\rm h}$ and $M_{\rm HI}$ for $M_{\rm h}\sim10^9$--$10^{13}\,M_{\odot}$, featuring an exponential cutoff at lower halo masses and a power law at higher halo masses:
\begin{equation}
    M_{\rm HI}(M_{\rm h}) = M_{0,\text{HI}} \left(\frac{M_{\rm h}/M_\odot}{M_{\rm{min}}}\right)^{\alpha_\text{HI}} \exp{\left[-\left(\frac{M_{\rm{min}}}{M_{\rm h}/M_\odot}\right)^{0.35}\right]}, 
    \label{paco_eq}
\end{equation}
where $M_{\rm{min}}$ is the cutoff mass (essentially in units of $M_\odot$), $M_{0,{\rm HI}}$ is the overall normalization and $\alpha_{\rm HI}$ is the power law index at high halo mass. \cite{paco_2018} determine values for these parameters at $z\in\{0,1,2,3,4,5\}$. While we are looking into halos of $z\sim6$--8, the relation evolves little at the highest redshifts where \cite{paco_2018} perform their fitting. Based on the $z=5$ parameter values, we use $\alpha_{\rm HI}$ = 0.74, and scale the other parameters down to 70\% of their values at $z=5$, so that $M_{0,\text{HI}} = 1.9 \times 10^9 M_{\odot}$ and $M_{\rm{min}}=2.0 \times 10^{10}$. \added{This downscaling is an approximate extrapolation of the redshift evolution of these parameters at $z\gtrsim4$.}

Recall that our goal is to express the \cii{} luminosity of a halo as a function of its mass $M_{\rm h}$. Since we have $M_{\rm HI}(M_{\rm h})$, the missing ingredient is the coefficient $A_\text{\cii{}}$ in~\autoref{alpha}. We calibrate this proportionality at a SFR of $\rm{SFR}^*\sim1\,M_\odot\,$yr$^{-1}$, not uncommon for FIREBox simulated galaxies at $z\sim6$.

Throughout this work we obtain averages of certain properties as a function of halo mass and redshift based on the model of~\cite{Behroozi13b,behroozi_2013}, which connects halo mass to stellar mass and SFR through empirical models matching outputs from cosmological simulations to observational constraints. Here we interpolate across outputs from this model to find the halo mass $M_{\rm h}^*$ corresponding to $\rm{SFR}^*$ at $z\sim6$, and subsequently the corresponding \hi{} mass $M_{\rm HI}^*\equiv M_{\rm HI}(M_{\rm h}^*)$ based on~\autoref{paco_eq}. We find $M_{\rm HI}^* \sim2\times 10^9\,M_\odot$. Per~\cite{liang2023},
\begin{equation}
    \frac{L_\text{\cii{}}}{L_{\odot}} \sim 10^7 \times \frac{\rm{SFR}}{M_{\odot}\text{\,yr}^{-1}}, 
    \label{CII_SFR}
\end{equation}
meaning that we should have $L_\text{\cii{}}(M_{\rm h}^*)\sim10^7\,L_{\odot}$. This sets the value of $A_\text{\cii{}}=0.005$, meaning that when considering only the proportionality of \cii{} luminosity with \hi{} mass, the relation between halo mass and \cii{} luminosity is
\begin{equation}
    L_\text{\cii{}}(M_{\rm h})/L_{\odot} = 0.005 \times M_{\rm HI}(M_{\rm h})/M_{\odot},
\end{equation}
with $M_{\rm HI}(M_{\rm h})$ given by~\autoref{paco_eq}.


\subsubsection{$L_\text{\cii{}}$ as a function of metallicity}
\label{sec:cii_mhiz}

The gas-phase metallicity $Z$ determines the amount of carbon in neutral and ionized gas available to emit in \cii{} in the first place, and thus is fundamental to our model. The proportionality we have derived above effectively assumes a constant global gas-phase metallicity across all halos, but this is far from realistic as we expect significant evolution of metallicity on average with both halo mass and redshift. We would in fact like to calculate the metallicity $Z$ expected in each halo and use it as an additional variable that modulates the halo--\cii{} connection, so that
\begin{equation}
    L_\text{\cii{}}/L_{\odot} = \alpha_\text{\cii{}} \times M_{\rm{H_I}}/M_{\odot} \times Z/Z_\odot. 
    \label{new_metal}
\end{equation}

It is common to relate the metallicity of galaxies to their stellar masses $M_*$ and SFR, to the point where this relation is considered `fundamental' (at least out to some moderately high redshift) and dubbed the fundamental metallicity relation (FMR) and studied using a range of galaxy selections and parameterizations~\citep{Mannucci10,LL10,Hunt12,Cresci19,Curti2020}. We use the model of~\cite{behroozi_2013} again, using their fitting formula for average stellar mass $\avg{M_*}(M_{\rm h},z)$ as a function of halo mass and redshift, and interpolating their outputs for $\avg{\mathrm{SFR}}(M_{\rm h},z)$.

Note in particular that the stellar mass--halo mass (SMHM) relation parameterization is
\begin{equation}
    \log{(\avg{M_*} (M_{\rm h}))} = \log{(\epsilon M_1)} + f{\left[\log{\left(\frac{M_{\rm h}}{M_1}\right)}\right]} - f(0), 
    \label{stellarm}
\end{equation}
where $M_1$ is the characteristic halo mass, $\epsilon$ is a fitting factor and $f(x)$ is defined as 
\begin{equation}
    f(x) = -\log{(10^{\alpha x} + 1)} + \frac{\delta[\log{(1 + \exp{x})}]^{\gamma_f}}{1 + \exp(10^{-x})}, 
\end{equation}
where $\alpha$ is the power law index at low mass, $\gamma_f$ is a fitting index and $\delta$ is a fitting factor. Most of the fitting parameters include some redshift dependence, including
\begin{equation}
    \alpha = \alpha_0 + \alpha_1\left(\frac{1}{1+z}-1\right)\exp{\left(-\frac{4}{(1+z)^2}\right)},
\end{equation}
with best-fitting values of $\alpha_0=-1.412$ and $\alpha_1=0.731$. We refer the reader to~\cite{behroozi_2013} for further details on the other parameters.

To find metallicities, \cite{Heintz_2021} use the FMR parameterization of~\cite{Curti2020}, and we do the same:
\begin{equation}
    \tilde{Z}(M_*, \mathrm{SFR}) = \tilde{Z}_0 - \frac{\gamma}{\beta} \log\left[1 + \left(\frac{M_*}{M_0(\rm{SFR})}\right)^{-\beta}\right], 
    \label{fmr}
\end{equation}
where $M_0(\mathrm{SFR}) = 10^{m_0} \times \mathrm{SFR}^{m_1}$, with best-fit parameters $\tilde{Z}_0$ = 8.779, $m_0$ = 10.11, $m_1$ = 0.56, $\gamma$ = 0.31 and $\beta$ = 2.1. Here, the metallicity $\tilde{Z}=12+\log{(\mathrm{O/H})}$ is not in units of solar metallicity. \cite{Curti2020} note that the solar oxygen abundance corresponds to $\tilde{Z}=8.69$, meaning that
\begin{equation}
    \frac{Z(M_*,\mathrm{SFR})}{Z_\odot} = 10^{\tilde{Z}(M_*,\mathrm{SFR}) - 8.69}.\label{eq:zconv}
\end{equation}

All that remains is to recalibrate the proportionality between \cii{} luminosity and $M_{\rm HI}$, as we did while building the first version of our model. For the halo mass $M_{\rm h}^*$ corresponding to SFR$^*\sim1\,M_{\odot}\text{\,yr}^{-1}$ at $z\sim6$, we should expect a metallicity of $Z^*/Z_\odot\sim0.3$ based on the relations outlined above. One can already see that $\alpha_\text{\cii{}}\approx A_\text{\cii{}}/(Z^*/Z_\odot)\sim0.02$. To be more precise we have
\begin{equation}
    \frac{L_\text{\cii{}}(M_{\rm h},z)}{L_{\odot}} = 0.024 \times \frac{M_{\rm{HI}}(M_{\rm h})}{M_{\odot}} \times\frac{Z(M_{\rm h},z)}{Z_\odot}, 
\end{equation}
with $M_{\rm HI}(M_{\rm h})$ still given by~\autoref{paco_eq} but the expected metallicity value $Z(M_{\rm h},z)$ now given by Equations~\ref{fmr} and~\ref{eq:zconv} combined with $\avg{\mathrm{SFR}}(M_{\rm h},z)$ and $\avg{M_*}(M_{\rm h},z)$ based on~\cite{behroozi_2013}. Additionally, we assume a minimum halo mass of $M_{\rm h}=10^{10}\,M_\odot$ required for \cii{} emission, with the implication being that we do not expect lower mass halos to host carbon and/or diffuse gas in sufficient abundance for there to be appreciable cooling through \cii{} emission\added{ (cf., e.g., Table 2 of~\citealt{Chung_2020} and discussion in their Section 3.1)}.

\subsubsection{Adding log-normal scatter to metallicity}
\label{sec:scatter}
A halo model cannot capture all of the baryonic physics interior to each halo driving the \cii{} emission, but we can approximate such physics as a stochastic process affecting the \cii{} luminosity. In this case, we assume that the carbon abundances are most likely to be driven away from the expected value, and randomly scatter the metallicity $Z(M_{\rm h},z)$ by a log-normal distribution with standard deviation $\sigma_Z=0.4$ (in units of dex), which in turn will scatter the \cii{} luminosity with the same distribution since $L_\text{\cii{}} \propto Z$. We choose the fiducial value for $\sigma_Z$ based on previous work by~\cite{vizgan2022}, which showed that the log-linear correlation between \cii{} luminosity and \hi{} mass exhibits a log-space root-mean-square scatter of 0.39 dex. 

To add scatter to metallicity, we compute a positive scaling factor $\zeta_Z$ randomly drawn from a log-normal distribution with the following PDF:
\begin{equation}
    P(\zeta_Z;\sigma_Z) = \frac{1}{\sigma \zeta_Z \sqrt{2\pi}} \exp\left({-\frac{(\ln{\zeta_Z} - \mu)^2}{2\sigma^2}}\right), 
\end{equation}
where $\sigma = \sigma_Z\ln(10)$ and $\mu = -\sigma^2/2$, with $\mu$ chosen to preserve the linear mean value of $Z$ for a given halo mass and redshift. Then, from the expected metallicity value $Z(M_{\rm h},z)$ given halo mass and redshift, we may obtain a randomized metallicity simply by multiplying by $\zeta_Z\sim P(\zeta_Z;\sigma_Z)$. Therefore, when `painting' \cii{} luminosities onto dark matter halos in a cosmological simulation, we calculate $L_\text{\cii{}}$ from the \hi{} mass and the randomized metallicity $\zeta_ZZ$ found for each halo:
\begin{equation}
    \frac{L_\text{\cii{}}(M_{\rm h},z)}{L_{\odot}} = 0.024 \times \frac{M_{\rm HI}(M_{\rm h})}{M_{\odot}} \times\frac{\zeta_ZZ(M_{\rm h},z)}{Z_\odot}, \label{eq:lcii}
\end{equation}

\subsection{Noise model: fundamental sensitivities of Stage 1 and 2 LIM experiments}
\label{sec:noise}

Along with the signal, we generate forecasts of observational experiments. To do so, we convolve the pure signal by a Gaussian approximation to the main beam, and add Gaussian noise to the result. This work only aims to examine the \emph{fundamental} sensitivity that \cii{} LIM will be able to achieve in the near- to medium-term future, and so we omit considerations of either non-Gaussian sky noise (which will be the subject of significant near-future work from the first generation of \cii{} experiments) or interloper emission (rejection of which is already the subject of considerable theoretical efforts -- see, e.g.,~\citealt{Breysse15,LidzTaylor16,Cheng16,Sun18,Cheng20,Chen22}). 


To add Gaussian noise to our signal, we obtain the sensitivity per map voxel. We refer the reader to other works like Appendix B of~\cite{Chung_2020} and references therein for further details on noise-related quantities, but for mm-wave observations, the convention is to express sensitivities in terms of the noise-equivalent intensity (NEI):
\begin{equation}
    \rm{NEI} = \frac{\rm{NEFD}}{\Omega_{\rm{beam}}},
\end{equation}
where $\rm{NEFD}$ is the noise equivalent flux density (the noise level per instrumental beam) and $\Omega_{\rm{beam}}$ is the solid angle in the sky covered by the beam. Approximating the beam as having a Gaussian profile with full width at half maximum of $\theta_\text{FWHM}$, the beam solid angle is given by
\begin{equation}\Omega_\text{beam}=\theta_\text{FWHM}^2\cdot\pi/(4\ln{2})\approx1.122\theta_\text{FWHM}^2.\end{equation}

Given the NEI, the actual noise level per voxel is given by
\begin{equation}
    \sigma_{n} = \frac{\rm{NEI}}{\sqrt{t_{\rm{vox}} N_{\rm{feeds}}}},
\end{equation}
where $N_{\rm{feeds}}$ is the number of spectroscopic feeds, and $t_{\rm{vox}}$ is the average observing time per volume element (or voxel) of the survey volume. For a spectrometer with simultaneous coverage of its observing band, this would simply be the observing time per sky pixel, but a scanning spectrometer will incur a penalty based on the number of steps required to cover the entire observing band.

We will centre our observational forecasts on the CCAT facility, which will deliver the largest field of view out of all currently funded EoR \cii{} experimental projects. Based on the most recent specifications at $\nu\sim280$ GHz~\citep{CCAT_Prime_Collaboration_2022}, the two EoR-Spec instrument modules proposed for Prime-Cam target an angular resolution of $\theta_\text{FWHM}=48$ arcseconds, a frequency resolution of $\delta_\nu=2.8$ GHz (or $\nu/\delta_\nu\approx100$), and NEFD values of 64, 81, and 120 mJy s$^{1/2}$ per beam in first-, second-, and third-quartile weather (respectively) at a source elevation of 45$^\circ$. 

Like the~\cite{CCAT_Prime_Collaboration_2022} work, we assume that the survey spans two fields of 4 deg$^2$, and that EoR-Spec requires 42 steps to cover the full observing band. While the original~\cite{CCAT_Prime_Collaboration_2022} forecasts assume 6912 beams and 80\% yield (which would imply $\approx5530$ active detectors) and a weighted average NEI across the varied weather conditions, we assume 5040 active detectors on the EoR-Spec focal plane and a NEFD per beam of 72.5 mJy s$^{1/2}$ (a simple arithmetic mean of the first- and second-quartile NEFD values). These two differences end up mostly cancelling each other, as we assume slightly fewer active detectors but slightly better NEFD per beam, and we end up with sensitivity figures within 10\% of those obtained by~\cite{CCAT_Prime_Collaboration_2022}.

We consider two experimental scenarios.
\begin{itemize}
    \item The first scenario is a CCAT-like experiment, specifically one conforming to specifications of the proposed baseline CCAT deep spectroscopic survey (CCAT-DSS) of~\cite{CCAT_Prime_Collaboration_2022} with 2000 hours of observing time per 4 deg$^2$ field. Assuming a pixelization of the survey volume with a pixel size of $\theta_\text{FWHM}/\sqrt{8\ln{2}}=20''$, this gives $58$ seconds within each pencil beam of $20''\times20''$, but $t_\text{vox}\approx1.4$ seconds after accounting for the scanning spectrometer penalty (which is equivalent to assuming $t_\text{vox}=58$ seconds but $N_\text{feeds}=5040/42=120$).
    \item The second scenario is a future Stage 2 successor LIM experiment, with $4.8\times10^6$ spectrometer-hours per field (versus the $2000\times120=2.4\times10^5$ spectrometer-hours achieved with the baseline CCAT survey), or $t_\text{vox}N_\text{feeds}=38$ hours. Advancements in on-chip spectrometers and readout technologies should relieve some of the efficiency limitations of the current generation of spectrometers and enable $10^6$--$10^7$ spectrometer-hour surveys to deploy within the next decade~\citep{Karkare22}. The CCAT facility is fundamentally capable of hosting a Stage 2 mm-wave LIM experiment of this calibre, as Prime-Cam occupies only the central $\sim40\%$ of the total field of view delivered by the telescope, and future CCAT instruments could deploy many more spectrometer modules. We do not necessarily assume these modules will be scanning spectrometers like the EoR-Spec modules, but assume that the same NEFD is fundamentally achievable since the values we assume are already weather-limited.
\end{itemize}


\subsection{Generating cosmological line-intensity mock data cubes}
Having pulled out our halo model for \cii{} from the FIRE simulation analysis of~\cite{liang2023}, and defined experimental parameters for near-future \cii{} surveys, it only remains to simulate observations of the \cii{} signal. We use a suite of approximate cosmological simulations to generate catalogues of halos spanning the survey volume, and then generate 3D cubes of simulated observations by `painting' luminosities onto these halos.

\subsubsection{Cosmology to halos: the peak-patch method}

The peak-patch (or mass peak-patch) method is a model for identifying halos in the initial dark matter fluctuation field. Moving beyond the original peaks theory of~\cite{bond_1986} and the excursion set method of~\cite{Bond91}, the peak-patch method is a convolutional method, identifying mass peaks in the initial (Lagrangian) density field after processing through a hierarchy of smoothing filters at different scales, and finding the final properties of peaks as collapsed halos by solving for homogeneous ellipsoidal collapse and using second-order Lagrangian perturbation theory to resolve flow dynamics. First expounded and validated in the work of~\cite{bond_1996,BM96b,BM96c}, the peak-patch method has recently returned to work following re-implementation and re-validation by~\cite{Stein_2018}. The peak-patch method generates large ensembles of independent cosmological realizations with much greater ease than conventional N-body methods, making it suitable for signal forecasts as well as covariance estimation for novel statistics. This is particularly true in the context of LIM~\citep{Ihle19,COMAPDDE}, which demand large lightcone runs resolving relatively low-mass halos, since the peak-patch method can intrinsically evolve the initial mass peaks to a continuous range of final redshifts instead of having to stitch together snapshots at discrete time steps.

We use a suite of peak-patch simulations generated to probe the possible use of novel statistics to apply to CO LIM at EoR redshifts and previously detailed in a number of papers, including~\cite{COMAPDDE}. These simulations were designed to push to lower masses by a factor of several compared to the peak-patch simulations generated at $z\sim3$ for~\cite{Ihle19}, and so have a size of $(960$\,Mpc$)^3$ and a resolution of $5640^3$ cells. In principle, this should correspond to a minimum resolvable halo mass of $6.4\times10^9\,M_\odot$; after adjusting masses to match abundances predicted by the halo mass function (HMF) of~\cite{Tinker08} with empirical high-redshift corrections from~\cite{behroozi_2013}, we find our simulations should be complete down to $M_{\rm h}\approx1.2\times10^{10}\,M_\odot$, well matched to our assumed minimum emitter halo mass of $10^{10}\,M_\odot$. (Note that such adjustments are not strictly necessary particularly at lower $z$ than those considered here; \cite{Stein_2018} showed that peak-patch simulations agree out of the box with the~\cite{Tinker08} HMF within 10\% for $M_\text{h}\gtrsim10^{11}\,M_\odot$ and $z\leq2$.)

The simulations are lightcone runs with the centre of the box placed 8668\,Mpc away from the observer; although the minimum requirement for CO LIM simulations was to cover $z=5.8$--7.9, the lightcones actually cover a slightly wider interval of $z=5.6$--8.2. The box size then corresponds to a sky area of $6^\circ\times6^\circ$ at the highest redshifts accessed. As we only need to simulate patches of $2^\circ\times2^\circ$, it is theoretically possible to slice each simulation into nine semi-independent sub-volumes for a total of 2430 lightcones instead of 270. However, the present work does not focus on aspects like covariances of statistics, and 270 lightcones is already a sufficient ensemble to reliably identify average expectations and variances. We instead favour using wholly independent volumes and use the central 4 deg$^2$ of each lightcone.

\subsubsection{Halos to maps: mocking \texorpdfstring{\cii{}}{[C II]} clustering}

To consider how the \cii{} line-intensity field responds to the presence of our simulated halos, we use modified versions of \texttt{limlam\_mocker}\footnote{\url{https://github.com/georgestein/limlam_mocker}} and \texttt{lim}\footnote{\url{https://github.com/pcbreysse/lim}} codes that interface with each other to paint the halo--\cii{} `response function' of~\autoref{eq:lcii} onto our peak-patch lightcones, as well as appropriately handle convolution of the beam profile and addition of noise. The code is more generally applicable to other atomic and molecular line species, with the only prerequisite being an appropriate halo model for line emission.

The code uses the peak-patch simulation outputs -- which catalogue each halo's mass, Eulerian positions, comoving distances and redshift -- as well as a function to relate halo properties to line luminosities (in our case the $L_\text{\cii{}}(M_{\rm h},z)$ function as described in~\autoref{sec:secmodel}) and the observational parameters described in~\autoref{sec:noise}, including the survey extent in angular and frequency coordinates, the NEFD, $\theta_\text{FWHM}$, $N_\text{feeds}$, $\delta_\nu$, and the total survey time (from which the code derives $t_\text{vox}$).

For all halos in the simulation that fall within the specified observational volume, we compute the \cii{} luminosity of each halo, but any mass-luminosity function, for any atomic or molecular line, can be prescribed. We then bin all halos and thus their luminosities into discrete voxels based on the pixelization of the sky and the frequency channelization. The conversion of the total luminosity $L_{\text{\cii{}}, \rm{vox}}$ in each voxel to the voxel intensity $I_\text{\cii{}}$ is straightforward:
\begin{equation}
    I_\text{\cii{}} = \frac{c}{4\pi\nu_\text{rest}H(z)}\cdot\frac{L_{\text{\cii{}}, \rm{vox}}}{V_\text{vox}}
\end{equation}
where $V_{\rm{vox}}$ is the comoving volume of each voxel, and the prefactor $c/[4\pi\nu_\text{rest}H(z)]$ depends on the speed of light $c$, the rest-frame line emission frequency $\nu_\text{rest}$, and the Hubble parameter $H(z)$ at the emission redshift.

From our peak-patch lightcone simulations, we produce \cii{} intensity cubes of size $\theta_S=2$ deg along each angular coordinate -- i.e., right ascension (RA) and declination -- and $\nu_{\rm{obs}} = 270\pm20$\,GHz in frequency space. Based on the CCAT beam width of $\theta_{\rm{FWHM}}=48''$ or $(1/75)$\,deg, we require a pixel size equal to the scale of the corresponding Gaussian profile, $\theta_\text{FWHM}/\sqrt{8\ln{2}}$, yielding $(\theta_S/\theta_\text{FWHM})\sqrt{8 \ln{2}}=353$ pixels on each transverse side. Along the line of sight in frequency space, we have a bandwidth of 40 GHz and a spectral resolution $\nu/\delta_\nu\approx100$ or $\delta_\nu=2.8$\,GHz, which makes for 15 frequency channels.

The observing frequencies encompass \cii{} emission from $z=5.6$--6.6; at the central frequency we obtain $V_{\rm{vox}}\approx22$\,Mpc$^3$ and $c/[4\pi\nu_\text{rest}H(z)]\approx 7.2 \times 10^{-4}$\,Jy\,sr$^{-1}$\,Mpc$^3$\,$L_\odot^{-1}$. For limited purposes, we also produce intensity cubes at other redshifts, including at $\nu_\text{obs}=(226\pm14)$\,GHz using otherwise similar parameters.

We consider the raw \cii{} cube in some cases, but in most cases we will consider the cube after convolution with the angular Gaussian profile approximating the CCAT beam, and with the addition of Gaussian noise as described in~\autoref{sec:noise}.

\subsection{Sanity checks: luminosity functions and \texorpdfstring{\cii{}}{[C II]} deficits}

Before moving on to the primary results of our simulations, we show the results of a couple of sanity checks of the model, first against other $z\sim6$ \cii{} luminosity functions in the literature and then against the so-called \cii{} deficit. The latter is not so prominent in the range of luminosities primarily relevant to our work, but there is nonetheless weak downward evolution of $L_\text{\cii{}}/L_\text{IR}$ with increasing $L_\text{IR}$.

\begin{figure}
    \centering
    \includegraphics[width=0.96\linewidth,trim=0 8mm 0 0]{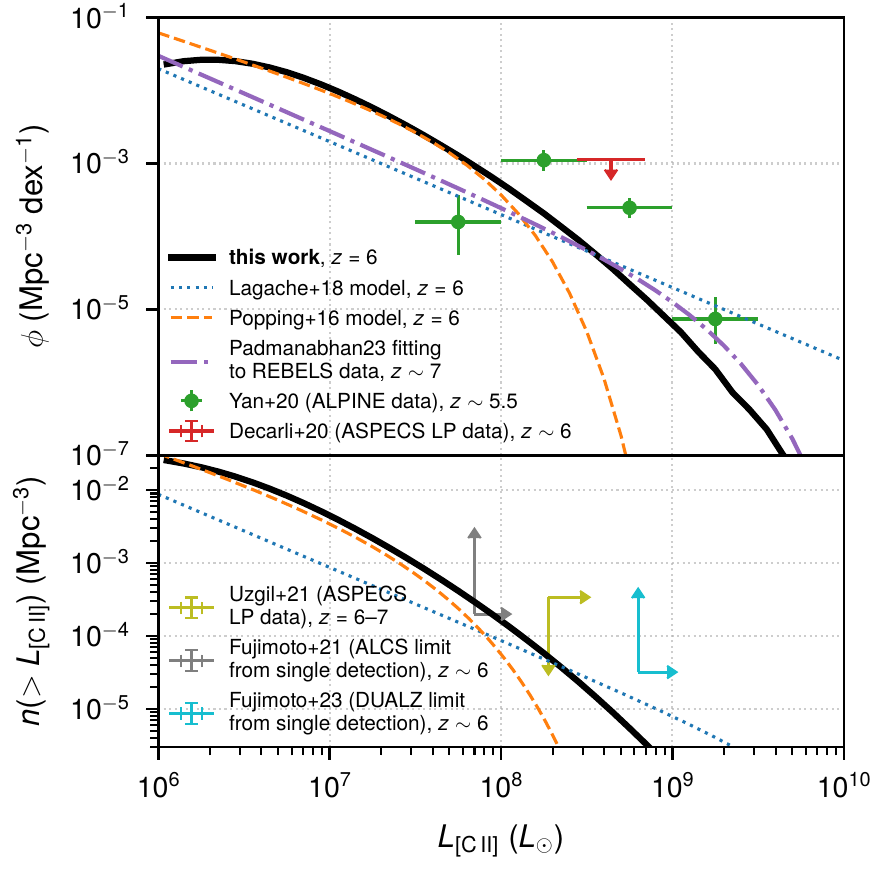}
    \caption{Comparison of this work's prediction for the $z=6$ \cii{} luminosity function $\phi=dn/d\log{(L_\text{\cii{}}/L_\odot)}$ (upper panel) and the cumulative number density $n(>L_\text{\cii{}})$ above different luminosity thresholds (lower panel), against a number of theoretical predictions~\citep{Popping16,Lagache_2018} and empirically derived constraints~\citep{Decarli20,ALPINELF,Padmanabhan23,Uzgil21,Fujimoto21,Fujimoto23}, as explained in the main text.}
    \label{fig:lf}
\end{figure}
As our peak-patch lightcones have halo masses adjusted to match the \cite{Tinker08}--\cite{behroozi_2013} form of the HMF, it is sufficient to use that HMF model to generate randomly drawn masses given some large cosmological volume, convert these masses to \cii{} luminosities (with randomly drawn metallicity values), and obtain the resulting luminosity function. We show this in~\autoref{fig:lf} alongside selected prior simulations and observations.

Our model output compares favourably to the prediction of~\cite{Popping16}, which uses a semi-analytic model to predict line luminosity functions at $z=0$--6; that work noted its difficulty in reproducing bright CO and \cii{} emitters at high redshifts, a difficulty which we do not share to the same extent. We also predict similar abundances of $L_\text{\cii{}}\sim10^8$--$10^9\,L_\odot$ objects as the model of~\cite{Lagache_2018}, although considerably fewer bright objects and somewhat more faint objects. (The comparison of~\cite{Bethermin22} in turn shows that the model of \cite{Chung_2020} predicts a luminosity function largely similar to that of~\cite{Lagache_2018}, as does their own model when using the \cite{Lagache_2018} SFR--\cii{} relation.)

We also show a similar level of consistency (i.e., within an order of magnitude) with observational constraints, which is remarkable given that we have not explicitly fit to them in great detail beyond the specific pivot point of $L_\text{\cii{}}\sim10^7\,L_\odot$ (which is in fact quite far from where most observational constraints lie).
\begin{itemize}
    \item The ASPECS LP survey was unable to go beyond upper limits on the luminosity function at $z\sim6$~\citep{Decarli20}; our simulated source abundances lie safely below the upper limits.
    \item Our simulated luminosity function lies systematically below the measurements of ALPINE~\citep{ALPINELF}, an ALMA survey of UV-selected galaxies. The target selection of ALPINE may be a source of bias, but the discrepancy against our model is certainly not as radical as against that of~\cite{Popping16} (which posited a similar explanation for inconsistency with a lower limit at $z\sim4$).
    \item The REBELS survey has also surveyed \cii{} emitters at $z\sim7$, but has not publicly released explicit luminosity function constraints; in its place, we compare the fitting of~\cite{Padmanabhan23} to the work in preparation against our model, and find consistency within a factor of 2--3 for $L_\text{\cii{}}\gtrsim10^8\,L_\odot$. For comparison, publicly discussed REBELS \cii{} emitters (e.g., in~\citealt{Ferrara22}) have luminosities of $10^{8.6}\,L_\odot$.
\end{itemize}

We also compare against constraints on the cumulative number density of \cii{} emitters above certain detection thresholds.
\begin{itemize}
    \item An alternative ASPECS LP analysis by~\cite{Uzgil21} derived upper limits on source abundances down to luminosities of $\sim10^8\,L_\odot$; we show their upper limit for sources at $z=6$--7 and demonstrate consistency with this limit. Although the original work of~\cite{Uzgil21} shows a number of other limits, the qualitative comparison against their other limits is the same.
    \item In addition, two surveys of lensing clusters have each yielded a single detection of a \cii{} emitter at $z\sim6$. ALCS~\citep{Fujimoto21} mapped strongly lensed regions across 33 different clusters, covering a total of 88 arcmin$^2$; DUALZ~\citep{Fujimoto23} specifically targets Abell 2744, in both a wide 24 arcmin$^2$ survey and a deep 4 arcmin$^2$ survey. As in the comparison against ALPINE data, our predictions lie systematically on the lower side of these limits. However, we caution for both ALCS and DUALZ that these constraints not only rely on a single detection in small effective survey areas of $\sim10^0$--$10^2$ arcmin$^2$, but also rely strongly on the mass model of the lensing cluster, both in the determination of the intrinsic luminosities that were or should have been detectable, and in the determination of the effective comoving volume surveyed.
\end{itemize}

For the $(L_\text{\cii{}}/L_\text{IR})$--$L_\text{IR}$ relation, we look to the simulated FIRE galaxies as a comparison point, rather than observational trends.\footnote{\added{We use results from~\cite{liang2023} where the prescribed cloud scale differs from that of their final paper, but the effects are only of order unity.}} Real-life resolved sources at $z>5$ are mostly exceptionally bright sources with $L_\text{IR}\gtrsim10^{12}\,L_\odot$ (and typically $L_\text{\cii{}}\gtrsim10^9\,L_\odot$), whereas it is almost entirely the complementary population that is most abundant and will dominate the large-scale clustering component of the \cii{} LIM signal.

To obtain the IR luminosity for our halos, we use the same IR--SFR conversion as~\cite{liang2023}:
\begin{equation}L_\text{IR}/L_\odot=1.36\times10^{10}\,\text{SFR}/(M_\odot\,\text{yr}^{-1}).\end{equation} This is the relation of \cite{Kennicutt98}, recalibrated for a~\cite{Kroupa02} initial mass function.

\begin{figure}
  \centering
    \begin{tikzpicture}
        \node () at (0,0) {\includegraphics[width=0.96\linewidth]{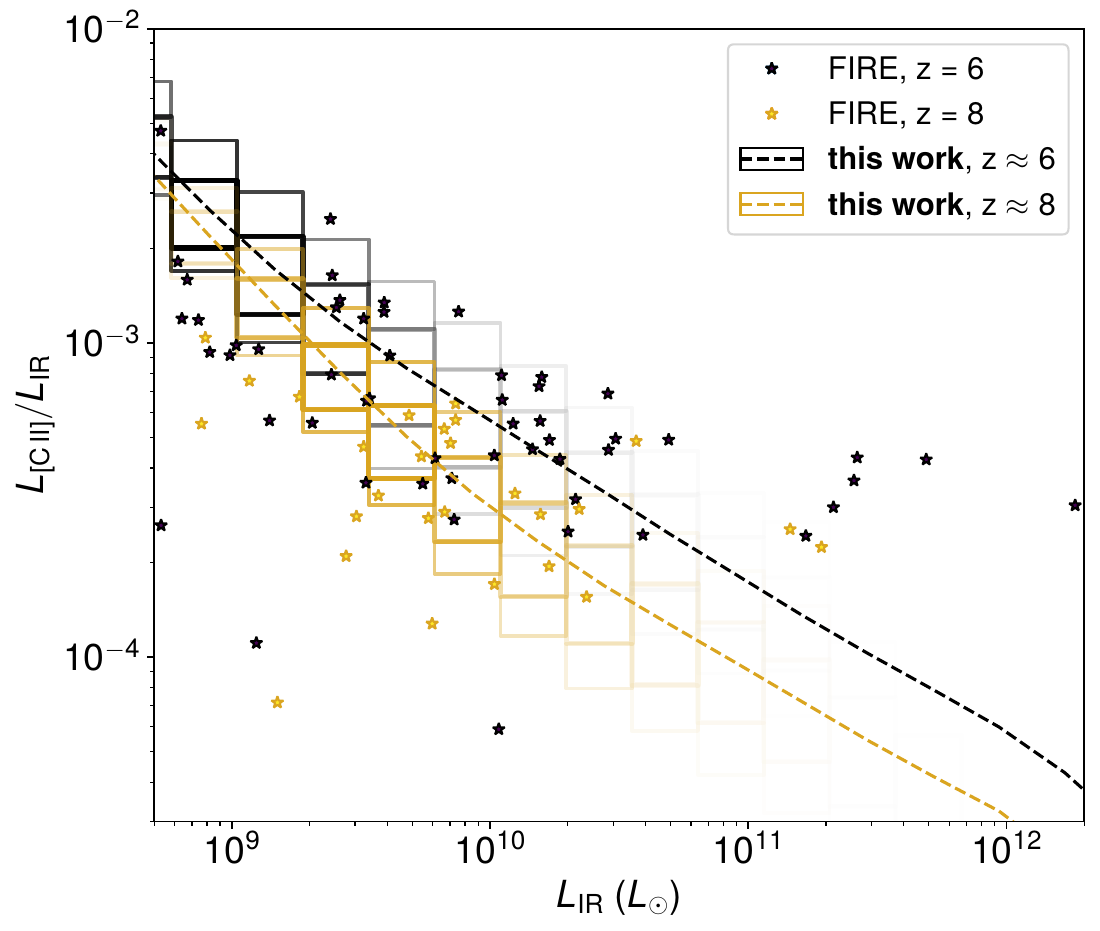}};
        \node () at (-0.86,3) {\fontfamily{qhv}\selectfont\large w/o metallicity dependence};
        \node () at (-0.86,2.586) {\fontfamily{qhv}\selectfont\large (incomplete model)};
    \end{tikzpicture}

    \vspace{-4.2mm}
    \begin{tikzpicture}
        \node () at (0,0) {\includegraphics[width=0.96\linewidth]{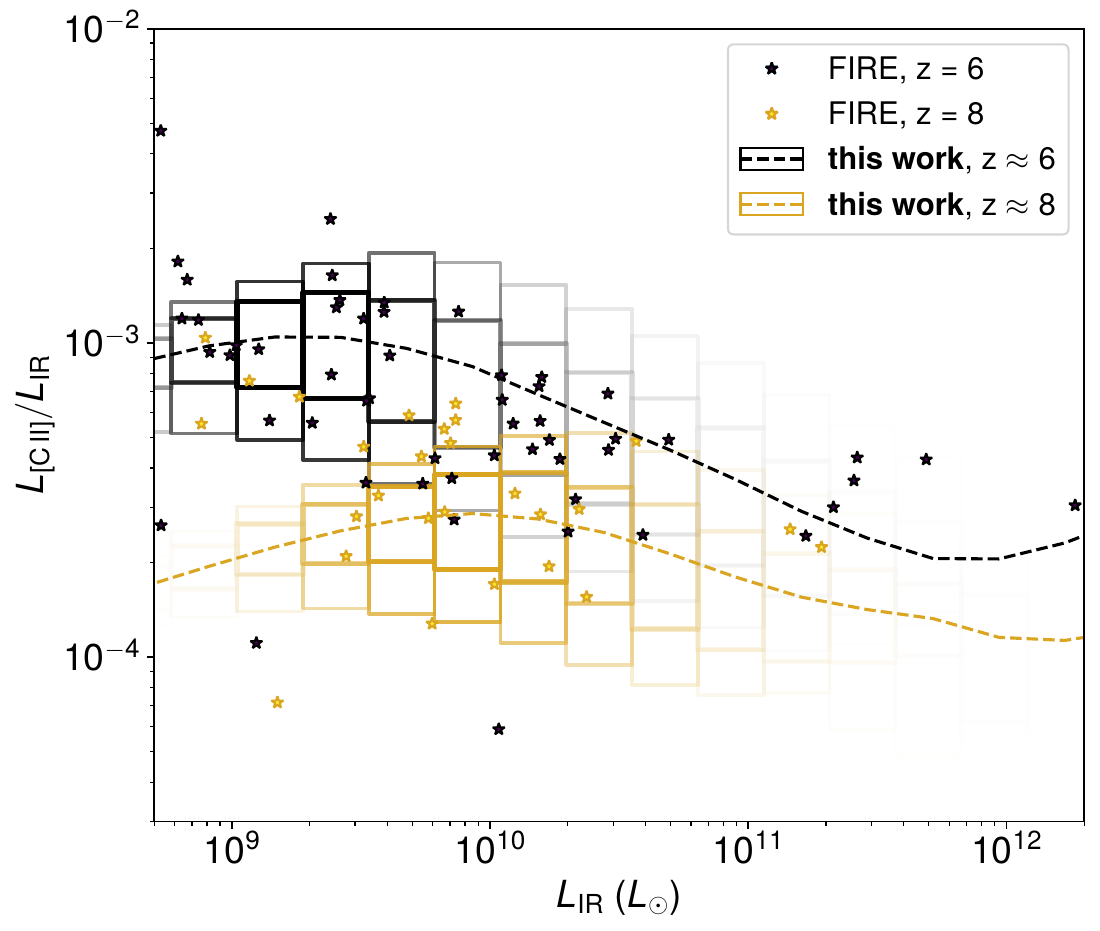}};
        \node () at (-0.86,3) {\bf\fontfamily{qhv}\selectfont\large w/ metallicity dependence};
        \node () at (-0.86,2.586) {\bf\fontfamily{qhv}\selectfont\large (full fiducial model)};
    \end{tikzpicture}
    \vspace{-4.2mm}
  \caption{Evolution of $L_\text{\cii{}}/L_\text{IR}$ with $L_\text{IR}$ with the incomplete model at the end of~\autoref{sec:cii_mhi} (upper panel), dependent only on \hi{} mass, and the full model described by~\autoref{eq:lcii} (lower panel) including metallicity. We show 68\% and 95\% intervals (unfilled rectangles, with more opaque bins being more abundant in our simulations) and the mean trend (dashed curves) in log-space luminosity bins of width $\approx0.25$ dex. Plotted alongside these are calculated values for simulated FIRE galaxies from similar redshifts.}\label{fig:deficit}
  \end{figure}
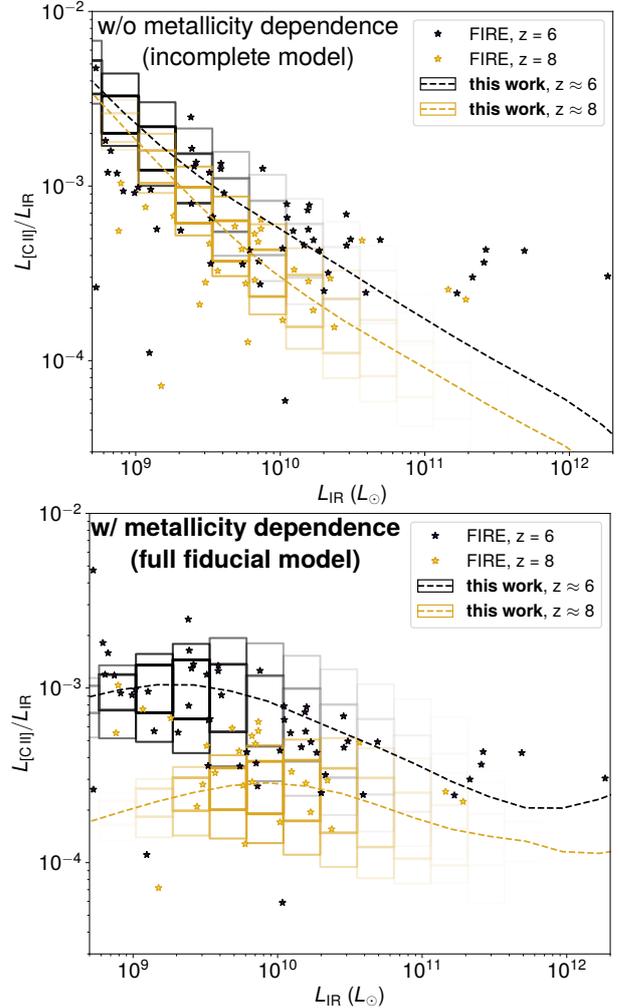

Recall that we formulated our final model in steps, and initially considered in~\autoref{sec:cii_mhi} a model with only explicit dependence on \hi{} mass (implying either fixed metallicity across all halos or lack of any metallicity-dependence). We show the $(L_\text{\cii{}}/L_\text{IR})$--$L_\text{IR}$ relation for this model in the upper panel of~\autoref{fig:deficit}. At high $L_\text{IR}$,~\cite{liang2023} note that $t_\text{dep}\sim M_\text{gas}/\mathrm{SFR}$ drives the so-called \cii{} deficit as there is less gas available to cool in \cii{} relative to the amount of star-forming activity. In the regime of $L_\text{IR}\lesssim10^{12}\,L_\odot$, however, the predicted evolution in $L_\text{\cii{}}/L_\text{IR}$ with $L_\text{IR}$ is far too steep when attempting to explain \cii{} emission with gas mass alone, and in particular over-predicts \cii{} emission in low-mass objects relative to the FIRE simulations.

The lower panel of~\autoref{fig:deficit} shows the $(L_\text{\cii{}}/L_\text{IR})$--$L_\text{IR}$ relation for our final model, using both \hi{} mass and metallicity to inform the \cii{} luminosity. As~\cite{liang2023} note, metallicity is clearly key in explaining a different kind of \cii{} deficit, which is the decline in $L_\text{\cii{}}/L_\text{IR}$ with increasing redshift at fixed $L_\text{IR}$. In general, less IR-luminous galaxies are less metallic and less abundant in carbon, which counteracts the increased relative availability of diffuse gas. The softened evolution of $L_\text{\cii{}}/L_\text{IR}$ with $L_\text{IR}$ is far more consistent with the simulated FIRE galaxy population. While the deficit is softened, we still predict a strong break in the $(L_\text{\cii{}}/L_\text{IR})$--$L_\text{IR}$ relation at $L_\text{IR}\gtrsim10^{10}\,L_\odot$ away from the na\"{i}ve expectation of constant $L_\text{\cii{}}/L_\text{IR}$.

It is possible that at the very highest IR luminosities shown in~\autoref{fig:deficit}, the dependence on metallicity over-compensates against the declining abundance of gas in general, and leads to an over-prediction of \cii{}.\added{ Inversely, metallicity-dependence suppresses our $z\approx8$ predictions at low and high $L_\text{IR}$ relative to FIRE predictions.} A crude parameterization of $f_\text{\cii{}}$ or a re-evaluation of the FMR at $z\gtrsim6$ would allow the model to capture the behaviour of the brightest\added{ and highest-redshift} sources, but this task is beyond the scope of the present work as \replaced{these}{$L_\text{IR}\gtrsim10^{12}\,L_\odot$} sources will be too rare to significantly affect the \cii{} LIM signal\added{ and as a precise replication of the FIRE outputs is not the primary focus of this work}.

\begin{figure*}
    \centering
    \includegraphics[width=\linewidth]{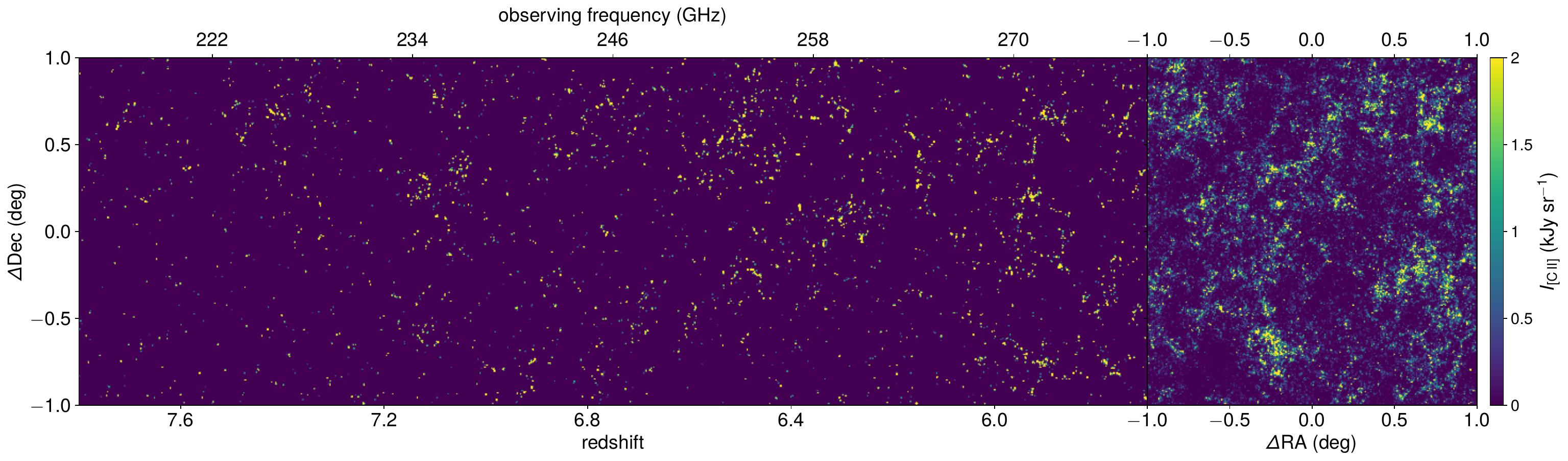}
    \caption{Slices taken from two \cii{} intensity cubes' simulations. One cube was generated with high frequency resolution ($\delta_\nu = 0.01$ GHz) to illustrate the evolution of the \cii{} signal along the frequency axis, while the other uses our base spectral resolution of $\delta_\nu = 2.8$ GHz to show a mock observation of \cii{}. \emph{Left panel:} slice of the $\delta_\nu = 0.01$ GHz cube at $\Delta\text{RA}\sim0$ deg with the frequency and redshift range indicated along the horizontal axes. Notice how the \cii{} power diminishes from lower to higher redshift. \emph{Right panel:} angular slice from the $\delta_\nu = 2.8$ GHz cube at $z \sim 6$ or $\nu_{\rm{obs}} \sim 270$ GHz.}
    \label{fig:prettyforecast}
\end{figure*}

\section{Simulation Results}
\label{sec:simres}
\subsection{Line-intensity maps and power spectra}

As previously outlined, we generate 270 lightcones of simulated \cii{} emission at $z\gtrsim6$. For most of this work, we will be focusing in particular on the part of the signal originating from $z=5.6$--$6.6$ (corresponding to $\nu_\text{obs}\in(250,290)$\,GHz). We plot the intensity map for a wider interval spanning $z\in(5.8,7.6)$ in~\autoref{fig:prettyforecast}, which shows the \cii{} emission tracing the large-scale clustering of our simulated dark matter halos, but with an intensity that falls off rather rapidly beyond $z\sim7$.

No actual experiment will perfectly recover the signal as shown in~\autoref{fig:prettyforecast}. As previously stated, the primary focus of LIM forecasts is the power spectrum, encapsulating the variance of the \cii{} line-intensity fluctuations at different scales, which should be detectable above noise. The power spectrum also lets us calculate a Wiener filter given the expected power spectrum and the actual noisy observation, thus constructing a minimum-variance linear estimator of the original \cii{} line-intensity map. We consider implementation details for the power spectrum and Wiener filter calculations in~\hyperref[sec:fft]{Appendix~\ref{sec:fft}}, and focus here on the results.

\begin{figure*}
    \centering\includegraphics[width=0.96\linewidth]{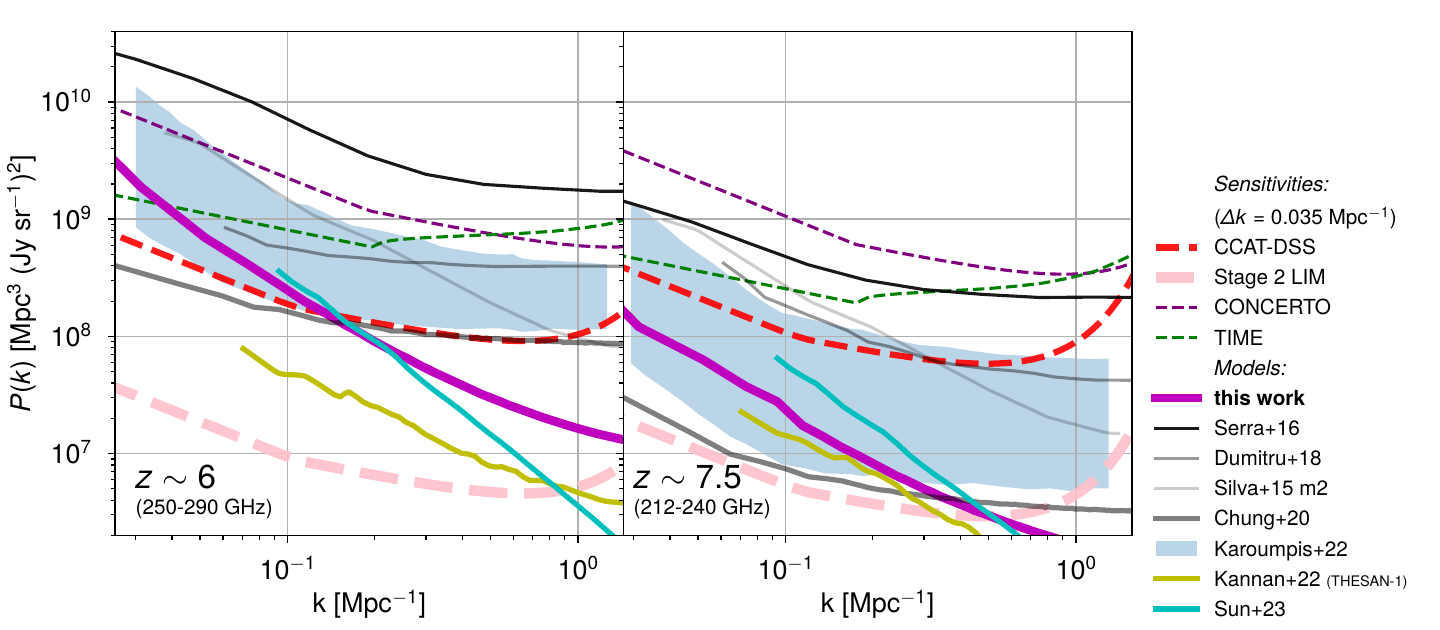}
    \caption{Models and sensitivities for the \cii{} intensity power spectrum at $z\sim6$ (left panel) and $z\sim7.5$ (right panel). Models are drawn either from simulations used in this work or from previous literature~\replaced{\citep{Silva15,Serra16,Dumitru18,Chung_2020,Karoumpis22}}{\citep{Silva15,Serra16,Dumitru18,Chung_2020,Karoumpis22,Kannan22,LIMFAST_II}}. Sensitivity forecasts for all surveys assume a mode count and noise level corresponding to the specifications outlined in either this work or in a previous work~\citep{Chung_2020} and wavenumber bins of width $\Delta k=0.035$\,Mpc$^{-1}$.}
    \label{fig:Pksens}
\end{figure*}

\autoref{fig:Pksens} shows the average power spectrum across all 270 lightcones, compared to both previous models~\replaced{\citep{Silva15,Serra16,Dumitru18,Chung_2020,Karoumpis22}}{\citep{Silva15,Serra16,Dumitru18,Chung_2020,Karoumpis22,Kannan22,LIMFAST_II}} and sensitivities expected from future \cii{} LIM experiments (see~\citealt{Chung_2020} for calculation details). At large scales, our predictions are similar to those of previous models if relatively pessimistic. However, our model is decidedly more pessimistic at small scales, indicating that we tend to predict significantly less shot noise compared to previous models, which is to say we do not predict a substantial population of rare bright \cii{} emitters.\added{ In this sense our predictions tie closest to those of~\cite{Kannan22} and~\cite{LIMFAST_II}.}

\begin{figure}
  \centering
    \includegraphics[width=\linewidth]{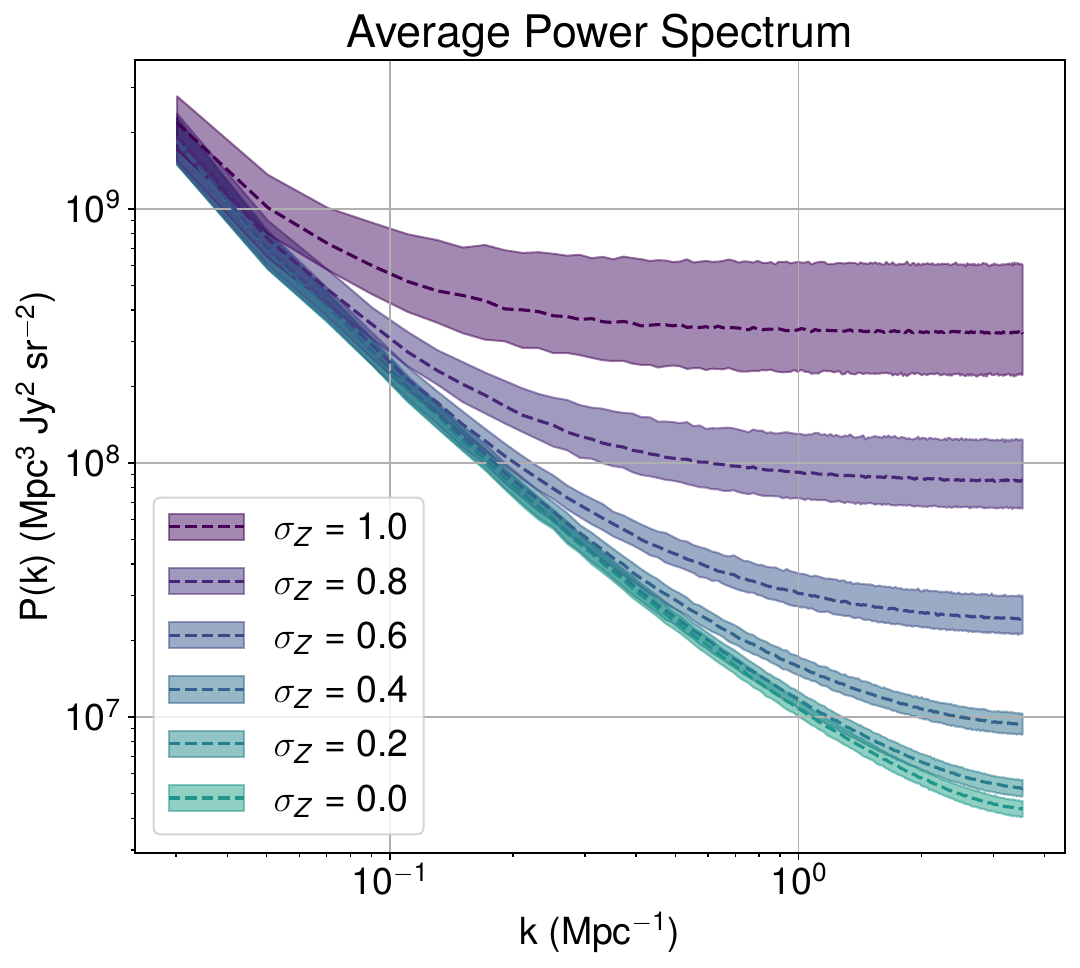}
  \caption{Average power spectrum of the \cii{} pure signal over our 270 halo lightcones. The dashed line indicates the mean and the shaded region indicates the $68\%$ interval about the median of the distribution the 270 lightcones' power spectra. Here, we display the average power spectrum for values of metallicity scatter index $\sigma_Z$ ranging from 0.0 to 1.0, which shows the gradual dominance of the shot noise across more scales with higher $\sigma_Z$.}
  \label{fig:pspecs}
\end{figure}

Greater stochasticity in \cii{} luminosity than the fiducial expectation could account for some of this difference, but not easily. \autoref{fig:pspecs} shows the \cii{} power spectrum for different amounts of log-normal scatter in metallicity ranging from 0.0 to 1.0 (in units of dex), where the fiducial value is $\sigma_Z=0.4$ dex. Higher scatter in metallicity introduces more random small scale fluctuations of the signal, and as a result, the clustering component of the power spectrum becomes subdominant to shot noise at lower $k$ for higher $\sigma_Z$. For an arbitrarily high level of scatter, the signal becomes entirely dominated by shot noise on all scales. However, the values required are extreme, with $\sigma_Z\gtrsim0.7$ dex required to achieve a level of shot noise relative to clustering similar to other model predictions. (As for why such values would be extreme relative to the fiducial $\sigma_Z=0.4$ dex, consider that a change in the standard deviation of a Gaussian variable from $\sigma=0.4$ to $\sigma=0.8$ would decrease the correlation coefficient against another variable by a factor of 2, all other covariances being held equal.)

In fact, the low level of shot noise arises in large part as a fundamental feature of our halo model: by its very motivation our model prescribes non-linearly suppressed \cii{} emission from more massive objects relative to their SFR, mimicking the so-called \cii{} deficit. By contrast, the other models use either a linear relation or mildly non-linear power law to obtain \cii{} luminosity from either SFR or IR luminosity. A key exception is \added{the prediction of~\cite{LIMFAST_II}, which uses the LIMFAST code~\citep{LIMFAST_I} to generate a model of \cii{} emission motivated by similar physical variables as in our model, albeit simulated at the molecular cloud level rather than at the dark matter halo level. Another exception exists in }one of the models of~\cite{Karoumpis22}, which works from the results of~\cite{Vallini15} to account for the effect of metallicity on the~\cii{} luminosity. But metallicity accounts for the \emph{redshift} evolution of the \cii{}--SFR relation given the metal-poor nature of early galaxies, while the primary driver of the \cii{} deficit \emph{at high luminosities} is the gas mass. We also see the reluctance of our model to prescribe very bright objects in the luminosity function shown in~\autoref{fig:lf}, especially in comparison to the model of~\cite{Lagache_2018}. All this suggests that the shape of the power spectrum and the relative balance of the clustering and shot noise components may be a key discriminator between competing physical pictures of \cii{} emission\replaced{ and}{. Although such pictures must include a wide range of processes that impact the halo--\cii{} connection at both low and high masses, the shape of the power spectrum may yet test different explanations} of the \cii{} deficit.

Prospects for detecting the power spectrum to this end are promising. When summed in quadrature across all $k\lesssim1\,$Mpc$^{-1}$, the total signal-to-noise ratio for the $z\sim6$ \cii{} power spectrum is $\approx\replaced{4}{5}\sigma$ for the baseline CCAT-DSS experiment. Although a detection at $z\sim7.5$ would be out of reach for CCAT-DSS, our Stage 2 LIM experimental parameters would be sufficient to achieve a $\approx\replaced{8}{11}\sigma$ overall detection of the $z\sim7.5$ \cii{} power spectrum, as well as an improved $\approx\replaced{36}{48}\sigma$ detection at $z\sim6$. The improvement would allow measurement of the shape of the \cii{} power spectrum beyond a simple initial detection, probing the contribution of populations of faint and bright galaxies to the total signal.

\begin{figure}
  \centering
    \includegraphics[width=\linewidth]{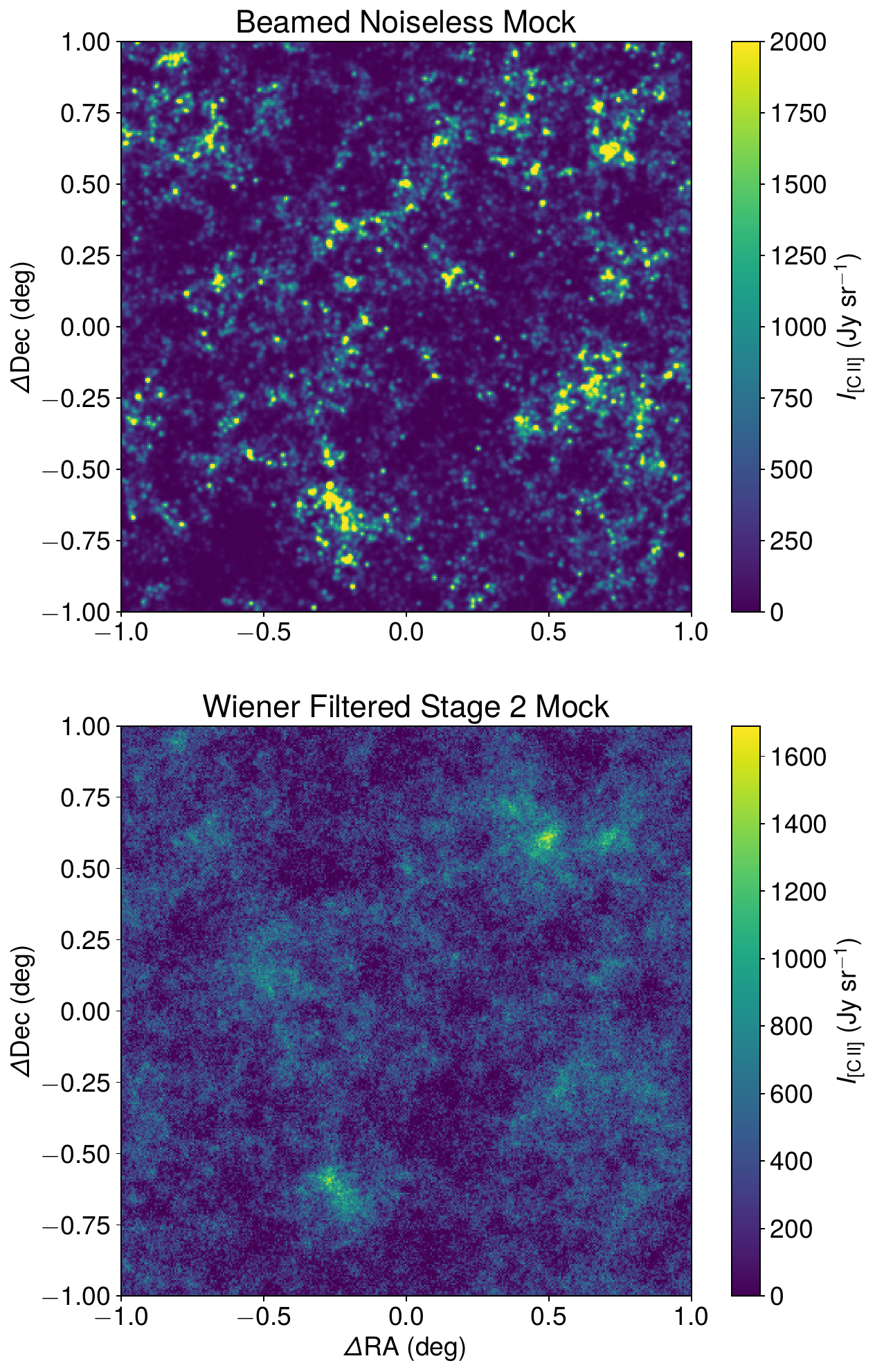}
\caption{Slices of a sample realization of input and recovered signal maps, both at $\nu_{\rm{obs}}\sim270$\,GHz or $z\sim6$. \emph{Upper panel:} Map of the \cii{} noiseless signal. Here, the map shown is convolved with a beam of 50 arcseconds to mock the CCAT beam width. Lower panel: Map of a simulated Wiener-filtered observation given parameters for a Stage 2 LIM survey.}
  \label{fig:forecasts}
  \end{figure}

A Wiener-filtered line intensity map provides a more intuitive illustration of what this detectability of the power spectrum means in terms of sensitivity to \cii{} intensity fluctuations at different scales. We process the raw line intensities from an example realization, as shown in the upper panel of \autoref{fig:forecasts}, through a Wiener filter given the Stage 2 forecast, ending up with the result in the lower panel of the same figure demonstrating recovery of prominent large-scale features (again, see~\hyperref[sec:fft]{Appendix~\ref{sec:fft}} for more information on the Wiener filter processing). 

\subsection{Stacking \texorpdfstring{\cii{}}{[C II]} on optical galaxies}
\label{sec:stacks}

We perform a mock stacking analysis to consider recoverability of the average \cii{} signal associated with locations of individual objects. Specifically, we assume that the halos in our simulations contain not only \cii{}-emitting gas, but also Lyman-break galaxy (LBG) populations that can be detected through photometric drop-out (as the Lyman break in the continuum flux falls between two photometric filters), and simulate stacks on a hypothetical narrow-band LBG survey that can access thousands of objects over the sky area of a single 2 deg$^2$ \cii{} LIM field.

We assume drop-out identification of LBGs at $z\in(5.9,6.1)$, and assign LBG counts to each halo in this redshift range. \cite{Harikane16} consider a halo occupation distribution model for a range of LBG populations detected in Hubble deep imagery and large-area Subaru/Hyper Suprime-Cam (HSC) data, including a $z\approx5.9$ LBG population detected as $i$-dropouts with threshold rest-frame UV aperture magnitude of 28.4, or absolute UV magnitude of $M_\text{UV}<-19.1$. Although this limiting magnitude is mostly driven by the Hubble data, which only span hundreds of square arcminutes, we assume that future ultra-deep photometric data will be capable of reaching this limit over areas on the scale of square degrees. This is not unreasonable based on the original HSC Subaru Strategic Program specifications~\citep{HSCSSP}, which specified a depth of $i\simeq28$ for the HSC-UltraDeep (3.5\,deg$^2$) layer; as of the third data release~\citep{HSCSSPDR3}, the HSC data had reached a depth of $i\simeq26.9$ in the Deep layer, with the UltraDeep layer being $\simeq0.8$ mag deeper.

We briefly recap the halo occupation parameterization of~\cite{Harikane16} parameterization here with minor changes to notation. The average number of LBGs in each halo is given as an assumed star-formation duty cycle $f_\text{duty}=0.6$ times the number of both central and satellite LBGs when active:
\begin{equation}
    \avg{N}(M_{\rm h}) = f_\text{duty}[\avg{N_\text{cen}}(M_{\rm h})+\avg{N_\text{sat}}(M_{\rm h})].
\end{equation}
The average central LBG count $\avg{N_\text{cen}}$ is given by a sigmoid which reaches 0.5 at some characteristic mass $M_\text{min,LBG}$ and has some width $\sigma_{\log{M}}$:
\begin{equation}
    \avg{N_\text{cen}}(M_{\rm h}) = \frac{1}{2}\left[1+\operatorname{erf}{\left(\frac{\log{M_{\rm h}/M_\text{min,LBG}}}{\sigma_{\log{M}}}\right)}\right],
\end{equation}
where for the $z\sim6$ LBG population, \cite{Harikane16} obtain $\log{(M_\text{min}/M_\odot)}={11.03}$ as the best-fitting value. The number of satellite LBGs is then this scaled by a power law:
\begin{equation}
    \avg{N_\text{sat}}(M_{\rm h}) = \avg{N_\text{cen}}(M_{\rm h})\left(\frac{M_{\rm h}-M_{0,\text{LBG}}}{M_1'}\right)^{\alpha_\text{LBG}},
\end{equation}
\cite{Harikane16} use results of prior empirical studies of halo occupation~\citep{Kravtsov04,Zheng05,Conroy06,MM15} to fix a number of parameters. This includes $\sigma_{\log{M}}=0.2$, $\alpha_\text{LBG}=1$, and
\begin{equation}
    \log{(M_{0,\text{LBG}}/M_\odot)} = 0.76\log{(M_1'/M_\odot)}+2.3,
\end{equation}
as well as
\begin{equation}
    \log{(M_1'/M_\odot)} = 1.18\log{(M_\text{min,LBG}/M_\odot)}-1.28.
\end{equation}

After obtaining $\avg{N}(M_{\rm h})$, we make a Poissonian draw to determine the `actual' number of LBGs hosted in each halo. This results in a typical count of $(4.3\pm0.4)\times10^3$ LBGs at $z\in(5.9,6.1)$ in each 2 deg$^2$ lightcone. The corresponding comoving number density of $\sim5\times10^{-4}$\,Mpc$^{-3}$ is lower than that found by~\cite{Harikane16} in data, but consistent with our own recalculation at $z=5.9$ using the HMF and the prescription for $\avg{N}(M_{\rm h})$, suggesting our results are likely consistent with their \emph{fitting} to the data (modulo cosmology differences).

We generate the actual mock stack on LBGs by cutting out a thumbnail from the \cii{} map centred on the host halo location, 50 pixels on each side ($\approx17'\times17'$), and then averaging the thumbnails weighted by the LBG count of all host halos.

\begin{figure}
  \centering
    \includegraphics[width=\linewidth]{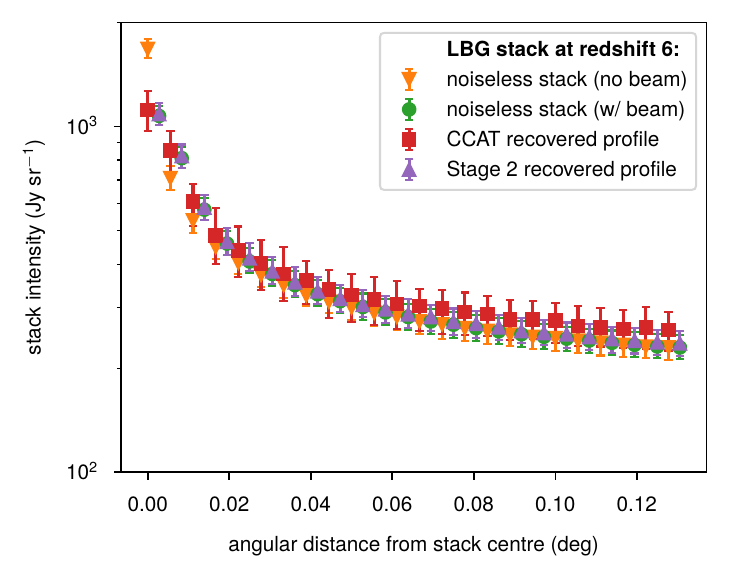}
  \caption{Radial average intensity profile of a simulated $z\sim6$ LBG stack for pure signal (both with and without a 48'' beam), a mock CCAT-DSS observation, and a mock Stage 2 LIM observation. The `noiseless stack (no beam)' and CCAT recovered profiles are slightly offset from their actual distance coordinates to show them more clearly against the other profiles.}
  \label{fig:stack_radpro}
  \end{figure}

We first show in~\autoref{fig:stack_radpro} the radial profile of the intensity of the stack away from the centre, as recovered from both noiseless simulations and realistic forecasts for \cii{} observations with CCAT-DSS and the Stage 2 concept. The profile appears to have two components, namely a more compact component localised near the stack centre corresponding to the \cii{} emission from the host halo, and a more diffuse component extending out to larger angular distances corresponding to emission from the surrounding large-scale structure. We find a slight upwards bias in the CCAT-DSS forecast for the second component at the level of 10\%, but otherwise the fundamental sensitivity of CCAT-DSS already allows for strong recovery of the radial profile of \cii{} emission around $z\sim6$ LBGs. With Stage 2 sensitivities the variance in the recovered profile is comparable to the corresponding variance for the noiseless stack, suggesting that instrumental noise in the \cii{} observation is not the dominant source of uncertainty in the radial profile in that scenario.

\begin{table}
    \centering
    \begin{tabular}{l|c|c|c}
         Parameter & Noiseless & CCAT-DSS & Stage 2 LIM \\\hline
         $I_{\rm D}$ (Jy\,sr$^{-1}$)&$\replaced{544\pm21}{669\pm26}$&$\replaced{538\pm24}{668\pm28}$&$\replaced{542\pm21}{667\pm27}$\\
         $r_0$ (px)&$0.39\pm0.22$&$0.\replaced{27}{32}\pm0.28$&$0.3\replaced{3}{6}\pm0.23$\\
         $\delta_{\rm D}$&$0.34\pm0.01$&$0.3\replaced{0}{1}\pm0.01$&$0.33\pm0.01$\\
         $I_{\rm G}$ (Jy\,sr$^{-1}$)&$\replaced{326\pm30}{401\pm37}$&$\replaced{352\pm42}{434\pm49}$&$\replaced{330\pm33}{408\pm41}$\\
         $\varrho$ (px$^2$)&$5.3\pm0.4$&$5.\replaced{5\pm0.6}{2\pm0.5}$&$5.3\pm0.5$\\ 
    \end{tabular}
    \caption{Recovered radial profile fitting parameters from stacks of noiseless, CCAT-DSS, and Stage 2 LIM mocks of a \cii{} observation on a $z\sim6$ LBG sample. Recall that 1 px in the map corresponds to an angular span of $20''$.}
    \label{tab:radpro}
\end{table}

We consider a phenomenological model of the radial profile as the sum of Gaussian and power-law profiles, with $r$ here being the radial distance in pixels (recalling that 1 px covers $20''$ along each angular dimension):
\begin{equation}
    \avg{I}(r) = I_{\rm D}\left(\frac{r_0+r}{1\,\text{px}}\right)^{-\delta_{\rm D}} + I_{\rm G}\exp{\left(-r^2/\varrho\right)}.\label{eq:stacktemplate}
\end{equation}
\autoref{tab:radpro} shows the recovered values of all five free parameters $\{I_{\rm D},r_0,\delta_{\rm D},I_{\rm G},\varrho\}$ from both the noiseless stack and the average expected stack from the CCAT-DSS and Stage 2 LIM forecasts, given respective errors. Note that while we na\"{i}vely expect $\varrho=2$ px$^2$ based on the beam profile being a Gaussian with scale of 1 px, the stacking procedure effectively convolves this with the pixel window function during the stack to widen the actual profile. Otherwise the recovered parameter values are generally consistent with what we see in~\autoref{fig:stack_radpro}, with the slight bias in the CCAT forecast for the large-scale component reflected in the lower recovered value of $\delta_{\rm D}$.

\begin{figure}
  \centering
    \includegraphics[width=\linewidth]{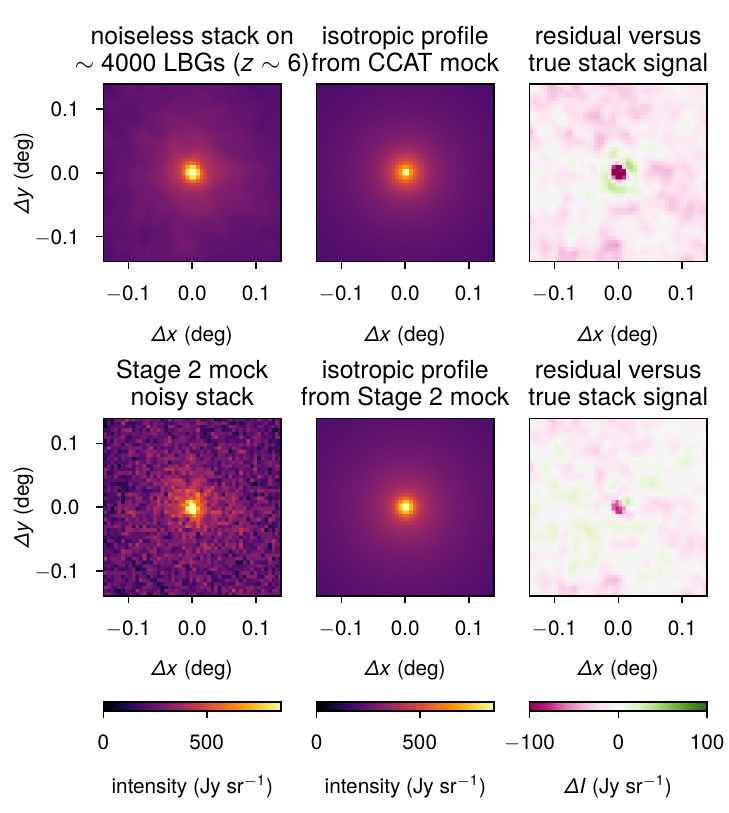}
  \caption{Simulated stacks on $\sim4000$ LBG locations from $z\in(5.9,6.1)$, including recovery for the CCAT and Stage 2 experimental scenarios \emph{from a single realization} of a 2 deg$^2$ field. \emph{Upper left:} The stack of the pure signal on the LBG locations, with only the Gaussian beam affecting the signal and no noise. \emph{Upper middle:} The best-fitting radial profile of the form described in~\autoref{eq:stacktemplate} recovered from stacking on a mock CCAT-DSS observation, re-projected into the stack thumbnail. \emph{Upper right:} Difference map between the recovered CCAT-DSS model thumbnail and the underlying true noiseless stack. \emph{Lower left:} The result of stacking on a mock Stage 2 observation. \emph{Lower middle:} Same as the upper middle panel, but with Stage 2 sensitivities. \emph{Lower right:} Same as the upper right panel but with Stage 2 sensitivities, demonstrating considerably reduced bias overall compared to the CCAT forecast in the upper right panel, but clearly showing some structure missed as a result of the isotropic nature of the simulated analysis.}
  \label{fig:stack}
  \end{figure}

For illustrative purposes, we also show a single realization of the stacking exercise in~\autoref{fig:stack}, including radial profiles recovered from that particular realization re-projected into a thumbnail. Note however that the Stage 2 LIM sensitivities allow very strong recovery of the \cii{} signal correlated with LBG locations without the need to isotropize the signal by averaging in radial bins. In fact, although recovery of the radial profile is already very good with CCAT-DSS sensitivities, an isotropized analysis even with Stage 2 sensitivities will miss out on the finer details of clustering of the \cii{} signal, as shown by the structure evident in the difference map between the recovered isotropized thumbnail and the true noiseless stack. Future work should explore the possibility of improving the information content of LIM stacks through anisotropic techniques like oriented stacking (see, e.g.,~\citealt{Lokken22,Lokken23}).

\subsection{Relative entropy in one-point statistics}
Returning to summary statistics of the \cii{} observation on its own, we now consider the ability of one-point statistics to distinguish the non-Gaussian signature of different parameters. 

Recall from~\autoref{sec:context} our definition of PRE per intensity bin:
\begin{equation}
    dS_\text{rel}(P\parallel Q) \equiv -P(I)\,\ln{\frac{P(I)}{Q(I)}}.
\end{equation}
Consider the `true' $P(I)$ to be our fiducial \cii{} model, and consider the alternate $Q(I)$ to be a version of the model with a change to some parameter $\lambda$. The PRE due to finding an observation with the fiducial value $\lambda_0$ of the parameter instead of the alternate value of $\lambda_0+\Delta\lambda$ is
\begin{equation}
    dS_\text{rel}(\Delta\lambda) = -P(I)\,\ln{\frac{P(I)}{Q(I;\lambda=\lambda_0+\Delta\lambda)}}.
\end{equation}
We may consider how this PRE varies with $\lambda$ (or equivalently its displacement $\Delta\lambda$ away from $\lambda_0$). Locally, this is simply given by the derivative of the PRE:
\begin{equation}
    \frac{\Delta(dS_\text{rel})}{\Delta\lambda} = -\frac{P(I)}{\Delta\lambda}\ln{\frac{P(I)}{Q(I;\lambda=\lambda_0+\Delta\lambda)}},
\end{equation}
or, taking $\Delta\lambda\to0$ while noting that $Q(I;\lambda=\lambda_0)=P(I)$,
\begin{equation}
    \frac{d(dS_\text{rel})}{d\lambda} = P(I)\frac{d}{d\lambda}[\ln{Q(I;\lambda)}].
\end{equation}
Therefore, the differential PRE incurred by a change in a given model parameter is simply the derivative of the log probability with respect to that parameter, weighted by the fiducial probability.

\begin{figure}
    \centering
    \includegraphics[width=0.965\linewidth,trim=-6mm 0 0 0]{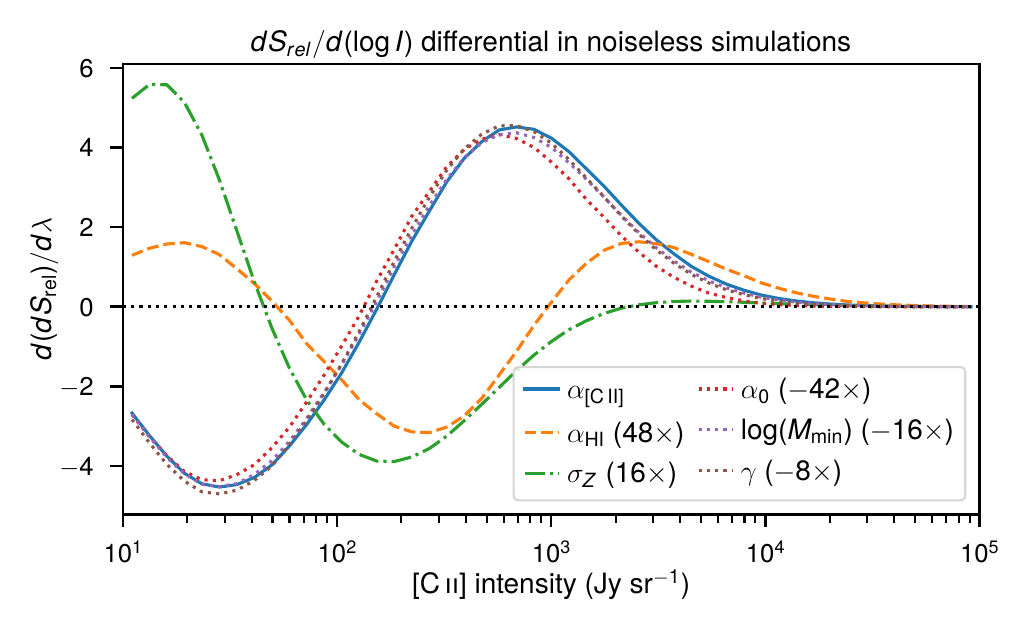}
    
    \includegraphics[width=0.97\linewidth]{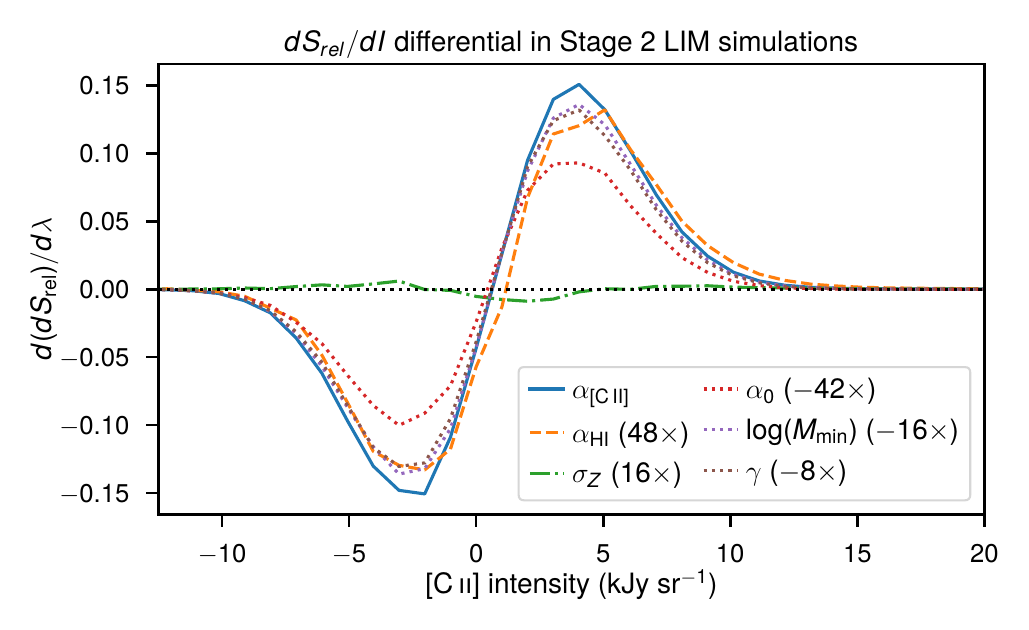}
    \caption{PRE derivatives with respect to a number of parameters (outlined in the main text), re-scaled as indicated in the legend, in either noiseless simulations of the \cii{} signal (upper panel) or simulations of a Stage 2 LIM observation of \cii{} (lower panel).}
    \label{fig:dSrel}
\end{figure}

We numerically estimate the PRE differential with respect to a number of parameters that we encounter in~\autoref{sec:secmodel}, namely:
\begin{itemize}
    \item $\alpha_\text{\cii{}}$, the overall normalization of $L_\text{\cii{}}(M_{\rm h},z)$ in~\autoref{eq:lcii}, with fiducial value of 0.024;
    \item $\sigma_Z$, the metallicity log-normal scatter in units of dex, with fiducial value of 0.4 dex;
    \item $\alpha_{\rm HI}$, the power law index of the $M_{\rm HI}(M_{\rm h})$ relation, with fiducial value of 0.74;
    \item $\log{M_\text{min}}$, the exponential cutoff mass scale in the $M_{\rm HI}(M_{\rm h})$ relation, with fiducial value of $10.3$;
    \item $\alpha_0$, controlling the power law index of the SMHM relation of~\cite{behroozi_2013}, with fiducial value of -1.412;
    \item and $\gamma$, a power law index in the FMR parameterization of~\cite{Curti2020}, with fiducial value of 0.31.
\end{itemize}

First, we consider the PRE in the noiseless signal simulation, before the introduction of either the angular beam or Gaussian noise. The upper panel of \autoref{fig:dSrel} shows the resulting PRE derivatives (per log-intensity bin) with respect to the six chosen parameters. Due to the significantly smaller fiducial value of $\alpha_\text{\cii{}}$ relative to those of the other parameters, we appropriately re-scale the PRE derivatives against the other parameters for easy comparison.

Since the PRE derivative is of the same sign as the derivative of the log-VID, the effect of increasing a given parameter $\lambda$ is to displace voxel counts towards `higher entropy' from bins of lower $d(dS_\text{rel})/d\lambda$ to bins of higher $d(dS_\text{rel})/d\lambda$. For instance, by increasing $\alpha_\text{\cii{}}$, we uniformly increase the brightness of the entire \cii{} intensity signal, thus pushing voxels from lower to higher intensities. For our fiducial model, this displacement appears to happen across a cross-over point of $\sim10^2\,$Jy\,sr$^{-1}$.

Unfortunately, the effect of many parameters is largely similar to that of $\alpha_\text{\cii{}}$, and their PRE derivatives appear mostly the same as $d(dS_\text{rel})/d\alpha_\text{\cii{}}$ after a rescaling and/or sign flip. Decreasing any one of $M_\text{min}$, $\alpha_0$, or $\gamma$ increases the brightness of the signal in a way that cannot be distinguished from increasing $\alpha_\text{\cii{}}$, at least not through the informational divergence in the one-point statistics of the map. In~\autoref{fig:dSrel}, we have specifically chosen rescalings that accentuate this degeneracy. (\hyperref[sec:dSrelalt]{Appendix~\ref{sec:dSrelalt}} shows the same derivatives with less drastic rescalings and against the natural log of the parameter value; the qualitative conclusions remain the same.)

We do however find some parameters whose effect appears distinct from that of $\alpha_\text{\cii{}}$. For example, increasing $\alpha_\text{HI}$ makes the $M_\text{HI}(M_{\rm h})$ relation steeper about the pivot point of $M_{\rm h}/M_\odot=M_\text{min}$, thus increasing \hi{} mass for halos with $M_{\rm h}/M_\odot\gtrsim M_\text{min}$ but decreasing it for halos with $M_{\rm h}/M_\odot\lesssim M_\text{min}$. This results in voxels with middling intensities becoming either brighter or dimmer, depending on the halos they contain. Similarly, increasing $\sigma_Z$ can either increase or decrease the metallicity, pushing voxels to both brighter and dimmer intensities.\footnote{\added{Note, of course, the relative simplicity of our model for stochasticity in metallicity and in \cii{} luminosity. Were we to incorporate additional parameters to allow $\sigma_Z$ to vary with halo mass, for example, this would alter the PRE signature of stochasticity, perhaps (but not necessarily) towards degeneracy with other parameters. Calculation of differential PRE templates for such additional parameters governing stochasticity should capture such added complexity.}} The characteristic intensity scales and effect sizes associated with these variables are different, meaning that $d(dS_\text{rel})/d\alpha_\text{HI}$ and $d(dS_\text{rel})/d\sigma_Z$ appear quite distinct from each other as well as from $d(dS_\text{rel})/d\alpha_\text{\cii{}}$.

However, much of this becomes insignificant against the effect of introducing the angular beam and Gaussian noise to the observation, which results in the PRE derivatives (per linear intensity bin) shown in the lower panel of~\autoref{fig:dSrel}. The PRE derivative with respect to $\alpha_\text{HI}$ is now indistinguishable in shape from that with respect to $\alpha_\text{\cii{}}$, while the PRE derivative with respect to $\sigma_Z$ still appears distinct but is significantly reduced in amplitude, suggesting its significantly reduced informational impact on the VID.

\subsubsection{Discussion point 1: observational distortions}

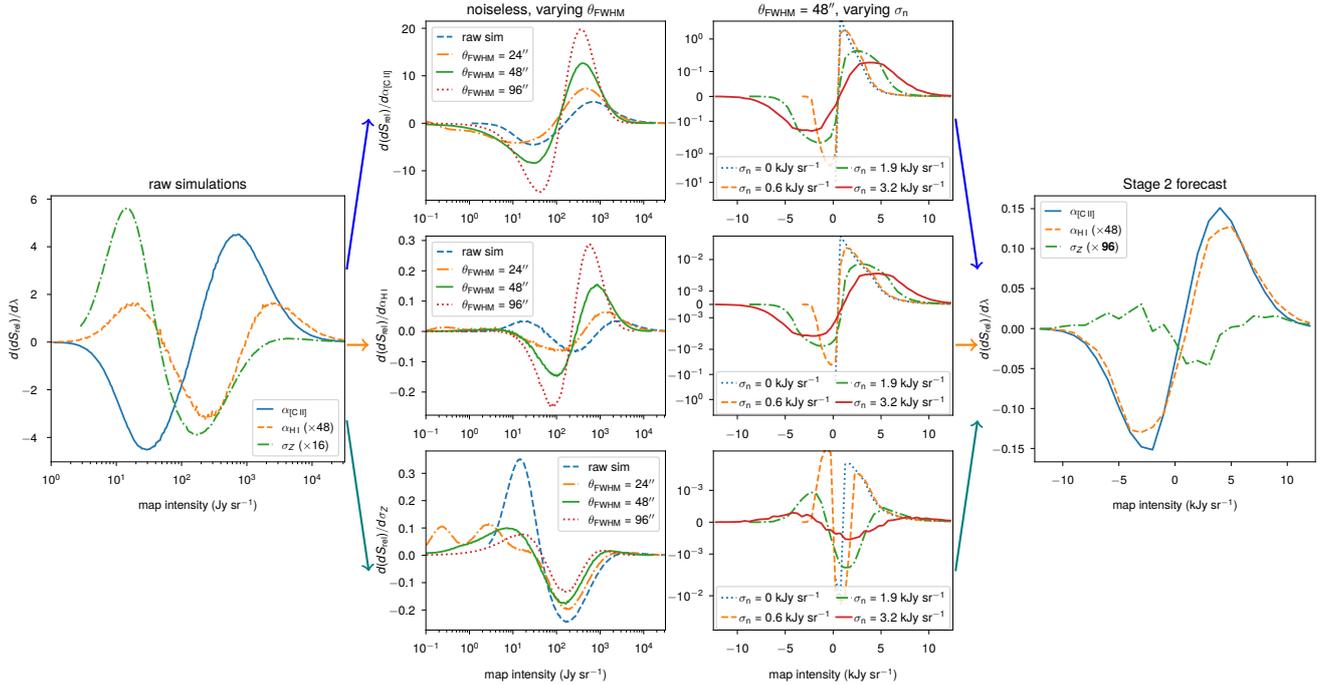
\begin{figure*}
    \centering
    \begin{tikzpicture}
        \node () at (0,0) {\includegraphics[width=0.45\linewidth]{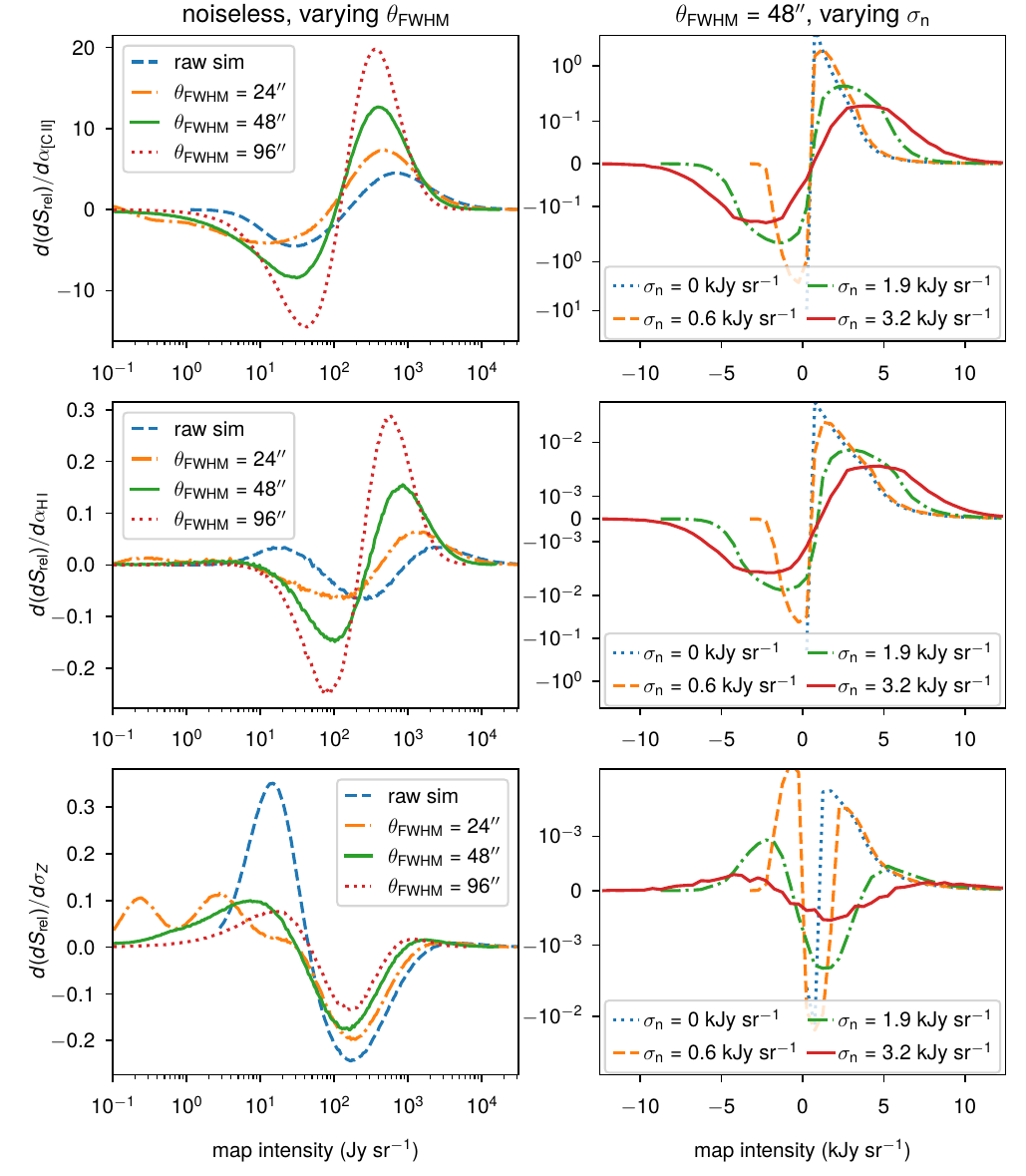}};
        \node () at (-6.386,0) {\includegraphics[width=0.2686\linewidth]{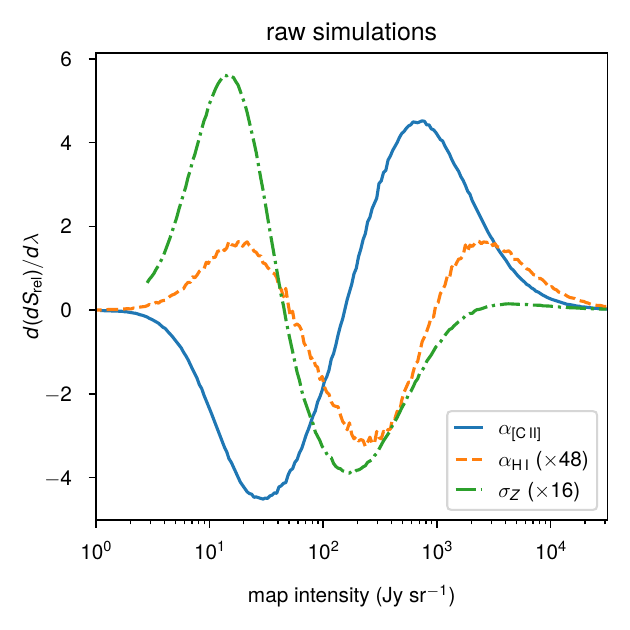}};
        \node () at (6.386,0) {\includegraphics[width=0.2686\linewidth]{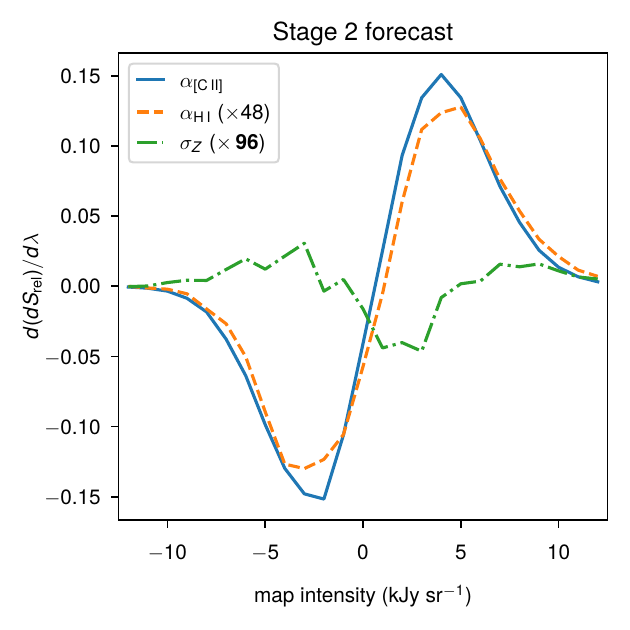}};
        \draw[->,thick,orange] (-4.15,0) -- (-3.86,0);
        \draw[->,thick,orange] (3.86,0) -- (4.15,0);
        \draw[->,thick,blue] (-4.15,1) -- (-3.86,3);
        \draw[->,thick,blue] (3.86,3) -- (4.15,1);
        \draw[->,thick,teal] (-4.15,-1) -- (-3.86,-3);
        \draw[->,thick,teal] (3.86,-3) -- (4.15,-1);
    \end{tikzpicture}
    \caption{Distortion of relative entropy derivatives due to observational effects. \emph{Leftmost panel:} Derivatives of pointwise relative entropy $dS_\text{rel}$ (evaluated in log intensity space) with respect to three parameters (scaled up as indicated in the legend): the $L_\text{\cii{}}$ proportionality constant $\alpha_\text{\cii{}}$, the $M_{\rm HI}(M_{\rm h})$ power law index $\alpha_{\rm HI}$, and the metallicity log-normal scatter $\sigma_Z$ (in units of dex). \emph{Inner left column of panels:} The same derivatives for all three parameters (from top to bottom panels), but with the introduction of a Gaussian beam, shown for several widths of the beam smoothing the signal and suppressing intensities. \emph{Inner right column of panels:} The derivatives of $dS_\text{rel}$ (now evaluated in linear intensity space) for the same three parameters (again from top to bottom panels), but with the fiducial Gaussian beam width of $\theta_\text{FWHM}=48''$, and with varying noise levels smoothing the actual intensity distribution. \emph{Rightmost panel:} The final pointwise relative entropy derivatives after accounting for both beam smoothing and Gaussian noise at the level of our Stage 2 experimental scenario. Note that as with the leftmost panel, these derivatives are also scaled up as indicated in the legend, and in particular the derivative with respect to $\sigma_Z$ is scaled up by a further factor of 6 compared to in the leftmost panel.}
    \label{fig:dSrel_distort}
\end{figure*}

It is clear that both smoothing of the signal by the beam and contamination from thermal noise cause significant loss and distortions in the informational sensitivity of the VID to our model parameters. To better understand how this occurs, we further vary the beam width $\theta_\text{FWHM}$ and noise scale $\sigma_{\rm n}$, and examine the resulting changes in the PRE derivative.

We summarise our findings in~\autoref{fig:dSrel_distort}. First, we consider the effect of the beam between the outer left panel and the inner left column of panels. Broadly speaking, the effect of convolving the field with a beam of finite size is a coarse pooling. This pooling dilutes the intensity in the brightest voxels across its neighbours, suppressing the heavy right tail that would originally be present in the VID and compressing the effect of each parameter to a narrower range of intensities. The effect is more complex at lower intensities, and so the distortion of the PRE derivative ultimately depends at some level on the clustering of the signal and the effect thereon of each model parameter.

Next, moving to the inner right column of panels of~\autoref{fig:dSrel_distort}, we consider the effect of introducing Gaussian noise, considering a few intermediate scenarios between the noiseless case and the Stage 2 forecast with $\sigma_{\rm n}=3.2\times10^3\,$Jy\,sr$^{-1}$. Inevitably, the PDF of the total intensity is the convolution of the individual signal and noise intensity PDFs, so introducing Gaussian noise means smoothing out the signal PDF by a Gaussian profile of size $\sigma_{\rm n}$. Given that we were plotting the PRE per log-intensity bin across a range of 1--$10^4$\,Jy\,sr$^{-1}$, the apparent result of convolution with the noise PDF is unsurprising. This is in fact the point where, as we move to considering the PRE derivative per linear intensity bin, the distinct signatures of some parameters homogenize, and overall the PRE derivatives clearly decrease significantly in magnitude as noise spreads the actual information content of the model parameter across more intensity bins. A lower value of $\sigma_{\rm n}\approx600$\,Jy\,sr$^{-1}$ -- effectively corresponding to a Stage 4 LIM survey -- better preserves the distinct nature of some PRE derivatives (e.g., against $\sigma_Z$), but the noise in a Stage 2 LIM experiment ends up having a more drastic effect.

By a combination of effective averaging of the signal and smoothing of the actual VID, the beam and Gaussian noise in our survey simulations tend to Gaussianize the observation and decrease the amount of information-theoretical divergence imparted to the VID by the change in any given model parameter.

\subsubsection{Discussion point 2: Paths forward for optimal information recovery from the VID}

While observational distortions do wipe away a significant amount of non-Gaussianity, it is possible to consider paths to extract the residual non-Gaussian signatures from the distorted VID. Detailed work on this front is beyond the scope of the present paper, but we will briefly discuss a few possible avenues.

Appropriate filtering or processing of the observational data could conceivably improve detectability of the distinct signatures of different parameters. For instance, the relative entropy analysis of~\cite{Lee23} used not the full simulated CIB map, but maps filtered in different bands of harmonic multipoles. Similarly, Fourier or wavelet analyses could reduce the impact of noise.

However, it seems difficult to extract an information-theoretical free lunch from the noisy LIM observation, at least on its own. Certainly the simplest noise-reducing filter, a rolling average, will ultimately tend to Gaussianize the signal even as it reduces the noise, which will be undesirable for certain signatures of skew or kurtosis. More advanced statistical methods could conceivably be better matched to the specific problem and perform better at isolating signal from noise in a non-linear, non-Gaussian fashion, but would likely require large amounts of training and thus be at least \emph{implicitly} constrained by prior expectations and other external information.

\begin{figure}
    \centering
    \includegraphics[width=0.96\linewidth]{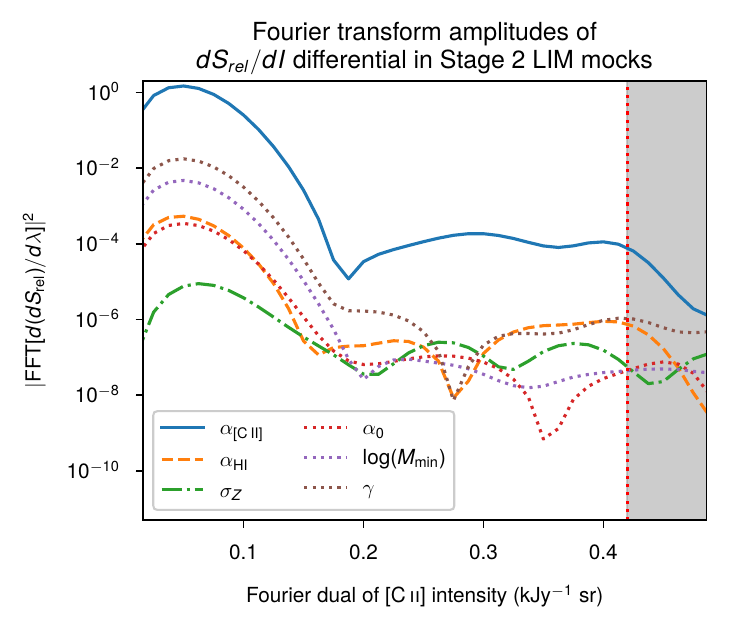}
    \caption{Power spectrum of the PRE derivatives for a simulated Stage 2 LIM survey (as shown in the lower panel of~\autoref{fig:dSrel}), representing the amplitude of the Fourier transform. The shaded area corresponds to values of the Fourier dual of voxel intensity that exceed $\pi/(2.355\sigma_{\rm n})$, where the noise of the Stage 2 LIM survey will likely dominate the variance of the characteristic function and derived quantities.}
    \label{fig:dSrel_fft}
\end{figure}

One of the best empirical paths forward likely involves \emph{explicitly} introducing external information and probing one-point statistics of cross-correlations or cross-correlations of one-point or higher-order statistics. Although the isotropic stacks of~\autoref{sec:stacks} do not necessarily reveal the clustering around the external source catalogue very prominently on casual inspection, an oriented stack that emphasises the filaments of the cosmic web may itself have one-point statistics that reveal non-Gaussian signatures of certain model variations. Similarly, the one-point statistics of a line-intensity map reconstruction through a non-linear extension of conventional techniques like linear covariance-based filtering~\citep{LIMLCB} could also access non-Gaussianities.

Another technique being developed for LIM applications is to deconvolve noise from signal through operations on the characteristic functions dual to the VID or counts-in-cells~\citep{Breysse19,Breysse23,COMAPDDE}. For reference, \autoref{fig:dSrel_fft} shows the spectral densities of the PRE derivatives. Purely as a qualitative heuristic, the spectral densities suggest that at smaller scales (i.e., higher values of the Fourier dual of the histogrammed intensity), the PRE does contain some small amount of information content imprinted in different ways by the distinct parameters of our model. Therefore, by leveraging the mutual one-point statistical information between a \cii{} LIM survey and a correlated observation, Fourier-space deconvolved distribution estimation (DDE;~\citealt{Breysse23}) can potentially access discriminatory information content by rejecting disjoint noise and systematics.

Finally, we note that accessing the information content of line-intensity maps is subject to further observational distortions that we have not accounted for -- e.g., scale-dependent sky noise, removal of interloper emission, and other mapmaking pipeline artefacts -- many of which will involve non-linear operations. Tailored simulations of specific experimental scenarios using large ensembles of numerical simulations of the kind we have considered here will thus continue to be critical to evaluation of information content, covariance estimation, and tests of inference pipelines.

\section{Conclusions}
\label{sec:conclus}
By now we have answered all of the questions we put forward in the~\href{sec:intro}{Introduction} to this paper:

\begin{itemize}\item \emph{What is the detectability of \cii{} emission in near- and far-future mm-wave LIM surveys, in either auto- or cross-correlation?} We expect a confident detection of the \cii{} power spectrum at $z\sim6$ with CCAT, followed by detection at $z\sim7.5$ with a Stage 2 LIM experiment. Stacking against a drop-out LBG survey can yield strong detections of the \cii{} signal near star-forming galaxies, as demonstrated with mock source densities corresponding to survey depths similar to those already achieved in existing surveys over multiple square degrees of sky.
\item \emph{How do different model parameters affect the voxel intensity distribution of the line-intensity fluctuations, as represented by the differential PRE?} The PRE derivatives of many parameters become apparently degenerate with that of a simple uniform scaling of the signal brightness, but some parameters like the metallicity scatter $\sigma_Z$ provide a distinct non-Gaussian signature that redistributes probability from middling intensities to both lower and higher intensities, or vice versa.
\item \emph{How do observational effects like the instrument response and thermal noise affect the observability of these distinct signatures in relative entropy?} By Gaussianizing the signal and smoothing out the VID, resolution effects and thermal noise both blur the signature in the log-VID, suppressing the observable PRE differential significantly and/or rendering all parameters degenerate with the overall brightness of the \cii{} fluctuations.\end{itemize}

We only stand at the beginning of the larger endeavour of \cii{} LIM, which will continue to require significant logistical and theoretical effort for optimal analysis and interpretation. Future work should investigate techniques like oriented stacking or the Fourier-space DDE -- or indeed entirely novel statistical methods -- leveraging external cross-correlation or mutual information to isolate signal from interlopers and noise. As we stated towards the end of the previous section, such methods may even be key to accessing sufficient information content to discriminate between different kinds of astrophysical and even cosmological non-Gaussian signatures. This includes signatures of primordial \emph{intermittent} non-Gaussianities propagating through large-scale structure and galaxy formation, which we will examine in future work through the lens of \cii{} LIM among other probes (Carlson et al., in prep.). In turn, these techniques and non-Gaussianities will require further study through rigorous simulation of both signal and noise.

In this aspect, a full understanding of observational effects will be important, but equally important is a full understanding of the \cii{} signal, or at least as full an understanding as one can get at this juncture. This work represents the state of the art in terms of a halo model grounded on physical insights from small-scale simulations, but even detailed simulations of individual galaxies are only as good as their inputs. Testing such models with \cii{} LIM will be desirable in future; this may demand the use not of manual tailored analyses of a simulation set (as in~\citealt{liang2023}) but of outputs of statistical methods like emulation or unsupervised dimensionality reduction in order to generate predictions for cosmological \cii{} fluctuations from a wide range of galaxy-scale models. Doubtless our understanding of the halo--\cii{} connection will continue to evolve with further observations of high-redshift \cii{}, though for now constraints on \cii{} at $z\gtrsim6$ rely on serendipitous detections and specific selections. It may happen that the first detections of cosmological \cii{} through LIM, in either auto- or cross-correlation, will arrive at constraints free from such selection effects to inform further future LIM experiments.

\section*{Acknowledgements}
Research in Canada is supported by NSERC and CIFAR. Parts of these calculations were performed on the Niagara supercomputer at the SciNet HPC Consortium. SciNet is funded by: the Canada Foundation for Innovation under the auspices of Compute Canada; the Government of Ontario; Ontario Research Fund -- Research Excellence; and the University of Toronto. Other parts of these calculations were performed on high-performance computing nodes at CITA; the authors thank John Dubinski and MJ Huang for tireless maintenance of these resources.

\added{The authors thank Jason Sun for kind and constructive comments on the present manuscript during the time that it was in review, and for providing data for~\autoref{fig:Pksens}.

}PH acknowledges support from CITA over the course of this research, including through the CITA Summer Undergraduate Research Fellowship and an NSERC Undergraduate Student Research Award.

Thanks to Patrick Breysse and Clara Chung, whose work with JRB under the 2020 Summer Undergraduate Research Programme in astronomy and astrophysics at the University of Toronto acts as the foundation of the present work.

DTC is supported by a CITA/Dunlap Institute postdoctoral fellowship. The Dunlap Institute is funded through an endowment established by the David Dunlap family and the University of Toronto. The University of Toronto operates on the traditional land of the Huron-Wendat, the Seneca, and most recently, the Mississaugas of the Credit River. DTC also acknowledges support through the Vincent and Beatrice Tremaine Postdoctoral Fellowship at CITA.

\added{LL acknowledges financial support from the Swiss National Science Foundation (grant number P2ZHP2\_199729) and the University of Toronto Faculty of Arts and Science.}

\added{Finally, we thank the anonymous referee that reviewed the manuscript and provided useful comments.}

This research made use of Astropy,\footnote{http://www.astropy.org} a community-developed core Python package for astronomy \citep{astropy:2013,astropy:2018,astropy:2022}, as well as the NumPy \citep{numpy:2020} and SciPy \citep{scipy:2020} packages. This research also made use of NASA’s Astrophysics Data System Bibliographic Services.


\section*{Data Availability}

The data underlying this article will be shared on reasonable request to the corresponding author.

\bibliographystyle{mnras}
\bibliography{biblio}

\appendix
\section{Implementation details for power spectra and Wiener filter calculations}
\label{sec:fft}

In order to carry out operations involving the power spectrum, we project the grid of our mock data cubes from observation space into comoving space. Given our cosmology and central redshift, it is straightforward to convert the $x$, $y$ (in angular space), and $z$ (in frequency space) coordinates of each voxel in our cube into comoving transverse and longitudinal positions ($x_\text{co}$, $y_\text{co}$ and $z_\text{co}$). We consider distortions from the lightcone geometry to be negligible for our simulated sky area and redshift range. We can compute the total span ($\Delta x_\text{co}$, $\Delta y_\text{co}$ and $\Delta z_\text{co}$) and the comoving distance per voxel ($dx_\text{co}$, $dy_\text{co}$ and $dz_\text{co}$) along each axis. Given the signal $t(\mathbfit{r})$ as a function of voxel coordinates $\mathbfit{r} = (x_\text{co},y_\text{co},z_\text{co})$, the power spectrum is then given as a function of the dual wavevector $\mathbfit{k}$:
\begin{equation}
    P(\mathbfit{k}) = \frac{|\operatorname{FFT}(\mathbfit{k}) \cdot dx_\text{co}\,dy_\text{co} \,dz_\text{co}|^2}{|\Delta x_\text{co}\,\Delta y_\text{co}\,\Delta z_\text{co}|},
    \label{eq:Pt}
\end{equation}
with $\operatorname{FFT}(\mathbfit{k})$ denoting the 3-dimensional discrete Fourier transform (DFT) of the signal $t(\mathbfit{r})$.

In terms of positional indices $(n_x,n_y,n_z)$ running from $(0,0,0)$ to $(N_x-1,N_y-1,N_z-1)$ and wavevector indices $(K_x,K_y,K_z)$ running through the same range, the DFT under our convention (the convention used by NumPy) is
\begin{align}
    \operatorname{FFT}(\mathbfit{K};t(\mathbfit{r})) &= \sum_{n_x}\sum_{n_y}\sum_{n_z} t(n_x,n_y,n_z) \times\nonumber\\&\qquad\exp{\left[-2\pi i\left(\frac{n_xK_x}{N_x}+\frac{n_yK_y}{N_y}+\frac{n_xK_z}{N_z}\right)\right]},
\end{align}
with the triple sum being over all voxel positions $\mathbfit{r}$ described by $(n_x,n_y,n_z)$.

We map the wavevector indices $(K_x,K_y,K_z)$ back to the comoving wavevector $\mathbfit{k} = (k_x, k_y, k_z)$ dual to $\mathbfit{r}$ based on the total width $(\Delta x_\text{co}, \Delta y_\text{co}, \Delta z_\text{co})$ and sampling $(dx_\text{co}, dy_\text{co}, dz_\text{co})$ along each dimension. The factor of the voxel volume in~\autoref{eq:Pt} applied to $\operatorname{FFT}(\mathbfit{k})$ rescales the DFT to be an approximation of the continuous Fourier transform, appropriate for use in the power spectrum calculation.

At each point of our Fourier transform cube, we compute the wavenumber $k=\sqrt{k_x^2 + k_y^2 + k_z^2}$. We define a 1D $k$-grid between 0 and the maximum $k$ value across the 3D $\mathbfit{k}$-space. We obtain an average of the 3D power spectrum $P(\mathbfit{k})$ weighted by the number of Fourier modes $N_\text{m}(k)$ belonging to each $k$-bin, and thus retrieve the spherically averaged power spectrum of our data cube.

\begin{figure}
  \centering
    \includegraphics[width=\linewidth]{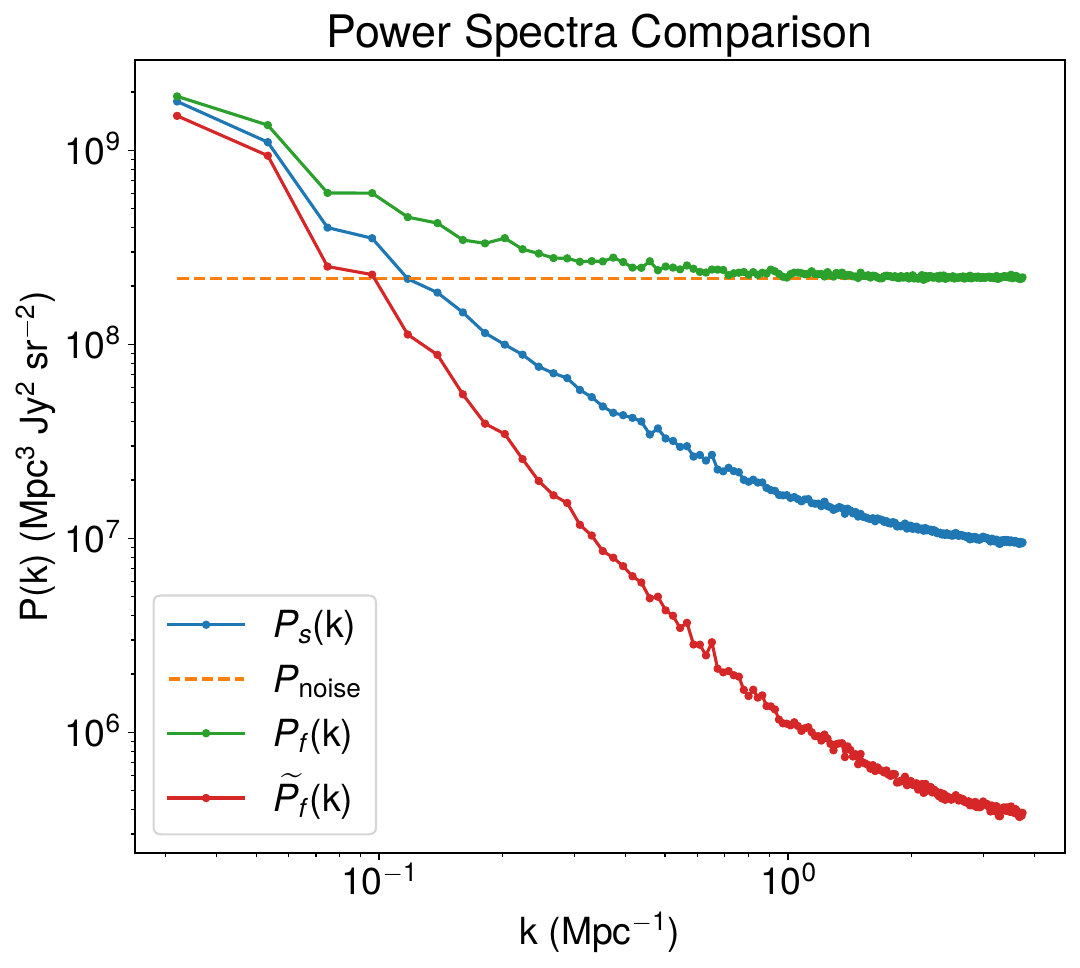}
  \caption{Comparison between power spectra of the noiseless \cii{} signal $P_s(k)$, of the Stage 2 observational \cii{} forecast $P_f(k)$, of the noise in the forecast $P_{\rm{noise}}$, and the Wiener-filtered Stage 2 \cii{} forecast $\widetilde{P}_f(k)$. Here, the analysis is carried out on the same lightcone used to generate the maps from Figure \ref{fig:forecasts}.}
  \label{fig:3pspecs}
  \end{figure}

Figure \ref{fig:3pspecs} shows the comparison between the 1D power spectrum of the pure signal $P_s(k)$ and the noise power spectrum amplitude $P_\text{noise}=\sigma_\text{n}^2V_\text{vox}$, as well as the power spectrum of the forecast observation $P_f(k)=P_s(k)+P_\text{noise}$. The sensitivities shown in~\autoref{fig:Pksens} are found effectively by taking $P_\text{noise}N_\text{m}^{-1/2}(k)$, while the signal to noise at each $k$ is found as $P_\text{s}(k)N_\text{m}^{1/2}(k)/(P_\text{s}(k)+P_\text{noise})$. Again, we refer the interested reader to~\cite{Chung_2020} for further details.

The Wiener filter is the optimal linear estimator for the signal given known signal and noise power spectra. Each Fourier mode is scaled by the following weighting function:
\begin{equation}
    W(k) = \frac{P_s(k)}{P_s(k) + P_\text{noise}}.
\end{equation}
Given the inverse discrete Fourier transform operation $\operatorname{IFFT}$, we can calculate the Wiener-filtered signal $\widetilde{t}(\mathbfit{r})$ from the forecast $T(\mathbfit{r})$ incorporating both signal $t(\mathbfit{r})$ and noise $\sim\mathcal{N}(0,\sigma_\text{n})$:
\begin{equation}
    \widetilde{t}(\mathbfit{r}) = \operatorname{IFFT}\left[\mathbfit{r}; W(|\mathbfit{k}|)\times\operatorname{FFT}(\mathbfit{k}; T(\mathbfit{r}))\right].
\end{equation}
(Note that our notation is intentionally slightly loose for simplicity, making implicit the mappings between discrete indices and continuous coordinates.) As~\autoref{fig:3pspecs} shows, this results in a filtered signal with a power spectrum given by
\begin{equation}
    \widetilde{P}_f(k)=W(k)^2P_f(k)=\frac{P_s^2(k)}{P_s(k)+P_\text{noise}},
\end{equation}
which approaches the original signal for signal-dominated $k$, and is heavily suppressed for noise-dominated $k$.

\section{PRE derivatives relative to model parameter fiducial values}
\label{sec:dSrelalt}
\begin{figure}
    \centering
    \includegraphics[width=0.965\linewidth,trim=-6mm 0 0 0]{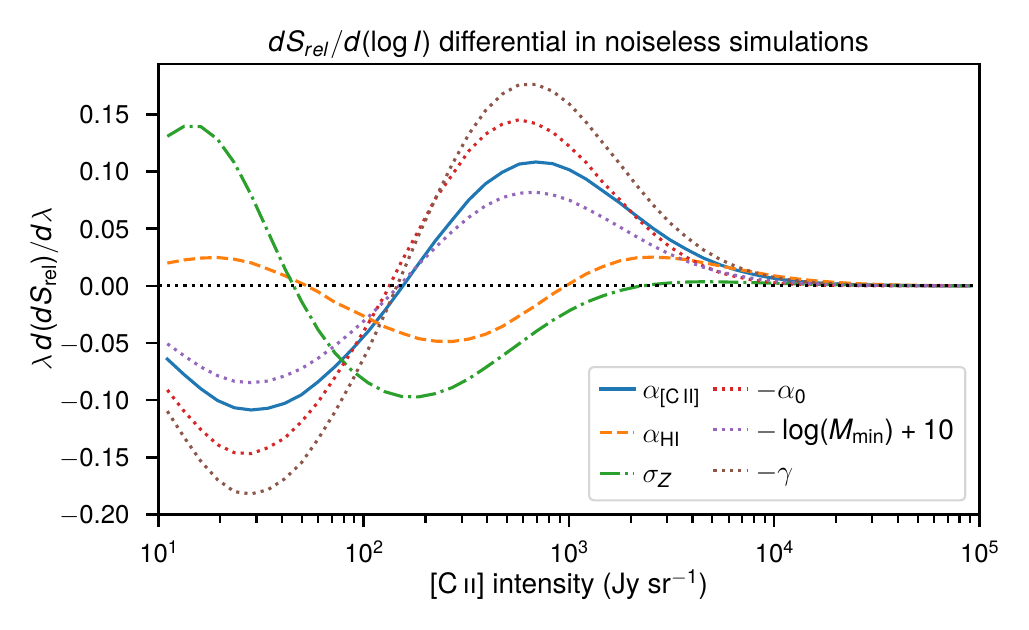}
    
    \includegraphics[width=0.97\linewidth]{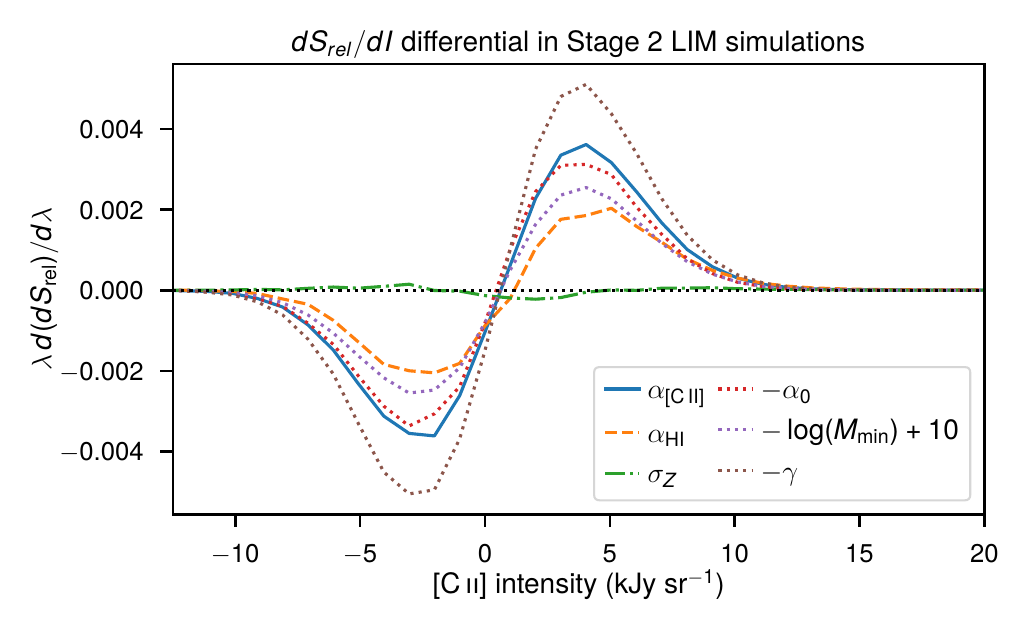}
    \caption{The same PRE derivatives shown in~\autoref{fig:dSrel}, but now against the natural log of the parameter values, and with different rescalings used. Half of the parameters are not rescaled or offset at all. For $\alpha_0$ and $\gamma$, we only apply a sign reversal; for $\log{M_\text{min}}$, we reverse sign and apply an additive offset of 10.}
    \label{fig:dSrel_alt}
\end{figure}

\autoref{fig:dSrel} in the main text shows the PRE derivatives with respect to the parameter values, and rescaled to emphasise a key degeneracy in the information content of multiple parameters. \autoref{fig:dSrel_alt} is an alternate visualization, albeit with very much the same effect. Here, instead of $d(dS_\text{rel})/d\lambda$ for each parameter $\lambda$ considered, we plot $\lambda\,d(dS_\text{rel})/d\lambda$. This is equal to the PRE derivative against the natural log of the parameter, or the PRE differential for a relative unit change in a parameter instead of an absolute unit change. We also only apply sign changes to better illustrate the same degeneracy observed between parameters. Otherwise we do not apply rescalings as aggressively as in~\autoref{fig:dSrel}, except for replacing $\log{M_\text{min}}$ with $\log{M_\text{min}}-10$.

\bsp	
\label{lastpage}
\end{document}